\documentclass[12pt]{article}
\usepackage[pdftex]{graphicx} %%GRAPHICX
\usepackage[light]{draftcopy}
\usepackage{epsfig}
\usepackage{comment}
\usepackage{latexsym}
\usepackage{color}
\newcommand{\mysquare}[0]{\raise-.2ex\hbox{{\Large$\Box$}}}
\def\lsim{\mathrel{\rlap {\raise.5ex\hbox{$ < $}}
{\lower.5ex\hbox{$\sim$}}}}
\def\gsim{\mathrel{\rlap {\raise.5ex\hbox{$ > $}}
{\lower.5ex\hbox{$\sim$}}}} \topmargin -1.5cm \textheight=22.5cm \textwidth=16.5cm
\catcode`\@=11
\newcount\hour
\newcount\minute
\newtoks\amorpm
\hour=\time\divide\hour by60 \minute=\time{\multiply\hour by60 \global\advance\minute by-\hour}
\edef\standardtime{{\ifnum\hour<12 \global\amorpm={am}%
        \else\global\amorpm={pm}\advance\hour by-12 \fi
        \ifnum\hour=0 \hour=12 \fi
        \number\hour:\ifnum\minute<10 0\fi\number\minute\the\amorpm}}
\edef\militarytime{\number\hour:\ifnum\minute<10 0\fi\number\minute}
\def\draftlabel#1{{\@bsphack\if@filesw {\let\thepage\relax
   \xdef\@gtempa{\write\@auxout{\string
      \newlabel{#1}{{\@currentlabel}{\thepage}}}}}\@gtempa
   \if@nobreak \ifvmode\nobreak\fi\fi\fi\@esphack}
        \gdef\@eqnlabel{#1}}
\def\@eqnlabel{}
\def\@vacuum{}
\def\draftmarginnote#1{\marginpar{\raggedright\scriptsize\tt#1}}
\def\draft{\oddsidemargin -.2truein
        \def\@oddfoot{\sl preliminary draft \hfil
        \rm\thepage\hfil\sl\today\quad\militarytime}
        \let\@evenfoot\@oddfoot \overfullrule 3pt
        \let\label=\draftlabel
        \let\marginnote=\draftmarginnote
   \def\@eqnnum{(\theequation)\rlap{\k

 ern\marginparsep\tt\@eqnlabel}%
\global\let\@eqnlabel\@vacuum}  }
\newcommand{\be}[0]{\begin{equation}}
\newcommand{\ee}[0]{\end{equation}}
\newcommand{\ba}[0]{\begin{eqnarray}}
\newcommand{\ea}[0]{\end{eqnarray}}

\def\bs{\begin{subequations}}
\def\es{\end{subequations}}

\def\thebibliography#1{%
\vskip 0.5cm \centerline{\bf \Large References}
\list{%
[\arabic{enumi}]}{\settowidth\labelwidth{[#1]} \leftmargin\labelwidth \advance\leftmargin\labelsep
\usecounter{enumi}}
\def\newblock{\hskip .11em plus .33em minus .07em}
\sloppy\clubpenalty4000\widowpenalty4000 \sfcode`\.=1000\relax}

\renewcommand{\theequation}{\arabic{section}.\arabic{equation}}

\renewcommand{\section}{\setcounter{equation}{0}\@startsection
{section}{1}{0mm}{-\baselineskip}{0.5\baselineskip} {\normalfont\Large\bfseries}}

\renewcommand{\subsection}{\@startsection
{subsection}{2}{0mm}{-\baselineskip}{0.5\baselineskip} {\normalfont\large\bfseries}}

\renewcommand{\subsubsection}{\@startsection
{subsubsection}{3}{0mm}{-\baselineskip}{0.5\baselineskip} {\normalfont\normalsize\slshape}}

\usepackage{amsmath}
\usepackage{amssymb,amsfonts}
\usepackage{graphicx}
\usepackage{cite}

\newcommand{\Z}{\mathbb{Z}}

\newcommand{\N}{{\cal N}}

\renewcommand{\and}{\mbox{and}}

\newcommand{\F}{{\cal F}}

\newcommand{\Th}[2]{\theta\left[^{#1}_{#2}\right]}

\begin{document}
%\verb|\usepackage{draftcopy}|\\
\begin{titlepage}
\begin{flushright}
LPTENS-10/05, January 2010 
\end{flushright}

%\vspace{2mm}

\begin{centering}
{\bf\Large Marginal Deformations of Vacua with\\ 
\vskip .2cm
Massive boson-fermion Degeneracy Symmetry$^\ast$}\\

\vspace{8mm}
 {\large Ioannis~Florakis$^1$, Costas~Kounnas$^1$ and Nicolaos~Toumbas$^2$ \\}

\vspace{8mm}

$^1$ Laboratoire de Physique Th\'eorique,
Ecole Normale Sup\'erieure,$^\dagger$ \\
24 rue Lhomond, F--75231 Paris cedex 05, France\\
\vspace{2mm}

\vspace{2mm}

$^2$ Department of Physics,
University of Cyprus \\
Nicosia 1678, Cyprus\\
\vspace{2mm}

{\it  florakis@lpt.ens.fr, kounnas@lpt.ens.fr, nick@ucy.ac.cy}

\vskip .1cm

 \vspace{3mm}

{\bf\Large Abstract}

\end{centering}
%\vspace{4mm}
\begin{quote}
Two-dimensional string vacua with Massive Spectrum boson-fermion Degeneracy Symmetry, $[MSDS]_{d=2}$, 
are explicitly constructed in Type II and Heterotic superstring theories. 
The study of their moduli space indicates the existence of large marginal deformations 
that connect continuously the initial $[MSDS]_{d=2}$ vacua to higher-dimensional conventional superstring vacua,
where spacetime supersymmetry is spontaneously broken by geometrical fluxes.
We find that the maximally symmetric, $[~{\rm Max}:\, MSDS~]_{d=2 }$, Type II vacuum, 
is in correspondence with the maximal, ${\cal N}$=8, $d$=4 ``gauged supergravity'', 
where the supergravity gauging is induced by the fluxes. This correspondence is extended to less symmetric cases 
where the initial \emph{MSDS} symmetry is reduced by orbifolds: \\ 
\centerline{$[~{\rm Z_{\rm orb} }:\, MSDS~]_{d=2 }$ $~~\longleftrightarrow~~ 
$ [~${\cal N}\le$8 :\, $SUGRA~]_{d=4,\, {\rm fluxes}}$ .}
We also exhibit and analyse thermal interpretations of some Euclidean versions of the models
and identify classes of \emph{MSDS} vacua that remain tachyon-free under arbitrary
marginal deformations about the extended symmetry point.
The connection between the two-dimensional \emph{MSDS} vacua and the resulting four-dimensional effective supergravities 
arises naturally within the context of an adiabatic cosmological evolution, 
where the very early Universe is conjectured to be described by an \emph{MSDS} vacuum, 
while at late cosmological times it is described by an effective $N=1$ supergravity theory 
with spontaneously broken supersymmetry.  
\noindent 
\end{quote}
\vspace{5pt} \vfill \hrule width 6.7cm \vskip.1mm{\small \small \small \noindent $^\ast$\ Research
partially supported by ANR (CNRS-USAR) contract 05-BLAN-0079-02.\\
$^\dagger$\ Unit{\'e} mixte  du CNRS et de l'Ecole Normale Sup{\'e}rieure associ\'ee \`a
l'Universit\'e Pierre et Marie Curie (Paris
6), UMR 8549.}

\end{titlepage}
\newpage
\setcounter{footnote}{0}
\renewcommand{\thefootnote}{\arabic{footnote}}
 \setlength{\baselineskip}{.7cm} \setlength{\parskip}{.2cm}

\setcounter{section}{0}

%%%%%%%%%%%%%%%%%%%%%%%%%%%%%%%%%%%%
%%%%%%%%%%%%%%%%%%%%%%%%%%%%%%%%%%%%

\section{Introduction}
The quest for vacua with spontaneously broken supersymmetry has 
always been one of the fundamental challenges for string phenomenology 
as well as for string cosmology. 
Whereas string models with spacetime supersymmetry 
arise quite naturally within the framework of superstring theory, 
naive attempts to break supersymmetry are usually plagued with tachyonic 
instabilities, similar in fact to the Hagedorn instabilities of string theory
at finite temperature.  

Ideally, one would like to start with a supersymmetric string compactification and 
induce a spontaneous breaking of  supersymmetry by turning on  non-trivial ``geometrical fluxes'' associated to a modulus. 
Vacua in the viscinity of the supersymmetric point would then acquire arbitrarily small supersymmetry breaking scale(s). However, if such a vacuum exists, it necessarily has to lie at the boundary\footnote{This is, of course, related to the well-known fact that the order parameter of the super-Higgs mechanism is fermionic, rather than bosonic. A very interesting exception that arises in theories with non-linear dilaton directions has been studied in \cite{HNO}.} of moduli space, where part 
(or all) of the internal space decompactifies.  A large class of vacua where the spontaneous breaking of supersymmetry arises via geometrical fluxes \cite{SSstringy,GeoFluxes} can be used to illustrate the main issues.
The geometrical fluxes introduce non-trivial correlations between $R$-symmetry charges $Q_I$ and the momentum and winding quantum numbers along non-trivial $S^1_I$ internal cycles.  As a result, the induced supersymmetry breaking scales are inversely proportional to the compactification radii, $M_I \sim 1/R_I$, and supersymmetry is recovered in the infinite-radius limit. On the other end, we often encounter tachyonic instabilities whenever $R \lesssim l_s$, which signal a non-trivial phase transition towards a new vacuum. 

One may also consider vacua where supersymmetry is spontaneously broken via geometrical fluxes at finite temperature $T$. Choosing, for instance, one supersymmetry breaking cycle with radius $R_1$, the induced supersymmetry breaking scale, $M = 1/(2\pi R_1)$, and the temperature scale, $T = 1/(2\pi R_0)$, 
provide a symmetric partition function, 
\be
Z(T,M)=Z(M,T), 
\ee
once the $R$-symmetry charge $Q$ is identified with the spacetime fermion number $F$. This is the simplest realization of a ``temperature/gravitino mass scale'' duality, exemplified by the Euclidean partition function $Z(T,M)$ \cite{ckpt}. It also illustrates how the existence of tachyons at small radii is related to the Hagedorn instabilities of string theory at high temperature.

In more general situations, where $Q_I \ne F$, the $T \leftrightarrow M$ exchange is not a symmetry of the theory 
but is replaced by other non-trivial duality relations involving the various supersymmetry breaking scales and the temperature.  In the context of non-perturbative string/string and M-theory dualities, we expect  a further non-perturbative generalization of such ``temperature/gravitino mass scale'' dualities, beyond the ``geometrical flux origin'' of supersymmetry breaking. We do not explore this generalization in the present work, even though we believe it to be of fundamental importance.  

An important implication of temperature/gravitino mass scale dualities is that {\it there may be several physical 
interpretations of the same Euclidean partition function} in models where supersymmetry is broken spontaneously 
by geometrical fluxes. For instance, when Euclidean time is identified with one of the non-compact
directions, the model describes a cold string vacuum, with the partition function determining the quantum effective potential, or the energy density of the vacuum. Alternatively, the Euclidean time  direction may be identified with one of the compact cycles $S_I^1$.  In this case the description is in terms of a thermal ensemble. The same Euclidean partition function determines the free energy and pressure of the thermal system.

The canonical ensemble corresponds to the case where the Euclidean time circle is completely factorized from the other cycles. The spin/statistics connection requires that the corresponding $R$-symmetry charge is identified
with the spacetime fermion number $F$. The canonical ensemble may then be deformed by turning on discrete ``gravito-magnetic'' fluxes associated to the graviphoton, $G_{0K}$, and axial vector, $B_{0K}$, gauge fields,
as in \cite{akpt}. Their presence refines the ensemble, and so chemical potentials for the graviphoton and axial vector charges appear. In Ref. \cite{akpt} thermal-like models of this type were constructed, where the Hagedorn instabilities are lifted for specific values of the chemical potentials. Moreover, the partition function was shown to be characterized by thermal duality symmetry: $R_0 \to \beta_c^2/R_0$ where $\beta_c \sim R_H$ (see also \cite{ACI,Chaud,DL}).    

Recently, significant progress towards the construction of tachyon-free, non supersymmetric  string vacua has been accomplished with the discovery of a novel \emph{Massive Spectrum boson-fermion Degeneracy Symmetry} 
(\emph{MSDS}), which is manifested at special points in the bulk of moduli space of a class of Heterotic and Type II orbifold compactifications to two (or even one) dimensions \cite{massivesusy,reducedMSDS}. 
At the \emph{MSDS} points all radii of the internal $T^8$ torus are at the fermionic point. Thus, the eight compact super-coordinates can be replaced by a set of $24$ left-moving and $24$ right-moving free fermions\cite{ABK,KLT}, with each set transforming in the adjoint representation of a semi-simple gauge group $H$ \cite{ABKW}. Whenever the boundary conditions respect the latter worldsheet symmetry group, spacetime gauge symmetry is enhanced to a non-abelian $H_L \times H_R$ local gauge group. The worldsheet degrees of freedom give rise to a new local 
superconformal algebra whose spectral-flow operator $Q_{MSDS}$  has the property of transforming bosonic into fermionic states at all massive levels, while leaving massless states unpaired \cite{massivesusy,reducedMSDS}. 
Because of the similarity with ordinary supersymmetry, we shall also make use of the term ``massive supersymmetry'', bearing in mind however, that the \emph{MSDS} symmetry is realized in terms of a different local algebra than ordinary supersymmetry. 
The degeneracy of states of mass $M$ is:
\be
n_B(M)-n_F(M) = \left\{ 
\begin{array}{c}
\neq 0~~\textrm{for}~ M=0\\
~0~~~~\textrm{for}~ M>0 \\
\end{array}\right. \, .
\ee
Therefore, the \emph{MSDS} vacua trivially satisfy the  condition of asymptotic supersymmetry     
\cite{KutSeiberg,Misaligned}, which in turn ensures the absence of (physical) tachyons from the spectrum. 

One motivation for studying these exotic vacua is the following. Within a string cosmological framework, it is quite natural to consider the possibility that the very early Universe arose as a hot compact space with characteristic curvature close to the string scale \cite{BV,massivesusy,reducedMSDS}. The underlying string dynamics may then drive three 
spatial directions to decompactify, with a large four-dimensional Universe emerging naturally. In the very early cosmological era, where the resulting description of spacetime is expected to be highly non-geometrical \cite{CosmoTopologyChange, ktt}, the full string theoretic degrees of freedom are relevant and must be properly taken into account. As the \emph{MSDS} vacua are non-singular, it was conjectured in \cite{massivesusy,reducedMSDS} that they are suitable candidates to describe this early stringy, non-geometrical era of the Universe. The high degree of symmetry present in these vacua opens a window in identifying and analysing the relevant string dynamics. 
Moreover, we will show in this paper that some Euclidean versions of the \emph{MSDS} models naturally describe thermal ensembles associated to $(4,0)$ supersymmetric models (or, more generally, $\mathcal{N}_L,\mathcal{N}_R\leq 4$), which are further deformed by discrete gravito-magnetic fluxes and are very similar to the tachyon-free thermal models of \cite{akpt}. The applicability of these thermal states in the context of hot string gas cosmology is currently under investigation \cite{BKPPT}.  
Previous work on string cosmology includes \cite{GV,HotstringsCosmo}.

As a first step in attacking the difficult problem of incorporating the backreaction due to thermal and quantum effects in \emph{MSDS} vacua, and in order to support the cosmological scenario outlined above, we analyse  
tree-level marginal deformations away from the enhanced symmetry points. The validity of such an adiabatic approximation requires the string coupling to be sufficiently small.  All marginal deformations are described in terms of worldsheet CFT perturbations of the current-current type $M_{IJ}J_I(z)\times\bar{J}_J(\bar{z})$,
where the $M_{IJ}$ describe the various moduli fields associated with the compactification\footnote{These fields are massless at tree-level, but may generically acquire mass from one-loop or higher genus corrections. The backreaction on the initially flat moduli fields due to quantum effects will be left for future investigation.}. As we deform away from the extended symmetry point, the \emph{MSDS} symmetry is broken spontaneously. 
Our goal is to identify classes of thermal and cold \emph{MSDS} vacua that remain tachyon-free under arbitrary marginal deformations, establishing tachyon-free trajectories that connect the two-dimensional \emph{MSDS} vacua to higher-dimensional supersymmetric ones via large marginal deformations.   

Below we outline the plan of the paper and a summary of our main results.

In Section \ref{review}, we briefly review the construction of the maximally symmetric \emph{MSDS} vacua \cite{massivesusy} and their $\Z_2^N$-orbifolds \cite{reducedMSDS}. We proceed in Section \ref{marginaldeformations} to study marginal deformations of the current-current type. 
 We show the existence of large marginal deformations interpolating between the maximally symmetric \emph{MSDS} vacua (and their orbifolds), and four-dimensional supersymmetic models, thus establishing a correspondence between the \emph{MSDS} space of vacua and four-dimensional gauged supergravities.  

In Section \ref{thermal}, we exhibit the thermal interpretation of various Euclidean versions of the models. The stability of the maximally symmetric vacua under arbitrary marginal deformations is analysed in Section \ref{stability}. We remark on the existence of marginal deformations of the maximally symmetric \emph{MSDS} models that can lead to tachyonic instabilities, and we show how the dangerous deformation directions can be projected out by considering asymmetric orbifold twists. Such orbifolds that preserve the \emph{MSDS} structure and produce vacua that remain tachyon-free under arbitrary marginal deformations are explicitly constructed in Section \ref{twisted}. We also identify conditions for general thermal versions of initially supersymmetric $(4,0)$ vacua to remain stable under arbitrary deformations of the dynamical moduli.    

In Section \ref{Hybrid}, we present a new class of  two-dimensional $(4,0)$ supersymmetric models, which will be referred to as `Hybrid' models throughout the paper. In these constructions the right-moving supersymmetries are
broken at the string scale and are replaced by the \emph{MSDS} structure. The left-moving supersymmetries are then broken spontaneously via a further asymmetric (freely acting) orbifold. The thermal partition function of the models exhibits thermal duality symmetry and is free of tachyonic instabilties.  

Finally, we present our conclusions and directions for future research.

%%%%%%%%%%%%%%%%%%%%%%%%%%%%%%%%%%%%
\section{Review of the \emph{MSDS} vacua}\label{review}

In this section we briefly review the maximally symmetric \emph{MSDS} vacua 
and their $\Z_2^N$ orbifold constructions. 
In their maximally symmetric version, the \emph{MSDS} vacua are constructed 
as two-dimensional compactifications $\mathcal{M}^2\times K$, 
where the compact space $K$ is described by a $\hat{c}=8$ superconformal field theory. 
The $8$ compact super-coordinates are described in terms of free worldsheet 
fermions \cite{ABK,ABKW,KLT}. 
This fermionization is possible at specific (``fermionic'') radii $R_i=\sqrt{\alpha'/2}$ 
for the compact worldsheet bosons. Throughout this paper we set $\alpha'=1$.

In Type II theories, the left-moving worldsheet degrees of freedom 
consist of the $2$ lightcone super-coordinates, 
the superconformal ghosts $(b,c),\, (\beta,\gamma)$ 
and the $8$ transverse super-coordinates $(\partial X^I, \chi^I)$ ($I=1,\ldots ,8$). 
The right-movers have a similar content in Type II theories, 
whereas in the Heterotic theory one adds $16$ extra complex fermions $\psi^A$, 
$A=1,\ldots,16$, as required by the cancellation of the right-moving 
conformal anomaly. 

Fermionizing (refs.\,\cite{ABKW, ABK}) the compact worldsheet bosons, 
we can express the abelian transverse currents $\partial X^I$ 
in terms of free worldsheet fermions $i\partial X^I(z) =y^I \omega^I(z)$. 
The fermions $\{\chi^I,y^I,\omega^I\}$ then realize a global affine algebra 
based on $G=\widehat{SO}(24)_{k=1}$. 
In the Heterotic theories, the right-movers can be fermionized without subtleties. 
For sectors with local worldsheet supersymmetry, however, as in Type II 
and the left-moving side of Heterotic theories, 
there are extra constraints arising from the fact that worldsheet supersymmetry 
must now be realized non-linearly among the free fermions:
$$
	\delta\psi^a\sim {f^a}_{bc}\psi^b\psi^c, ~~a=1,\ldots, 24.
$$
Imposing that this be a \emph{real} supersymmetry gauges the original symmetry $G$ 
down to a subgroup $H$, such that $G/H$ is a symmetric space. 
The free fermions then transform in the adjoint representation of the semi-simple Lie sub-algebra $H$, 
with $\dim{H}=24$. The local currents $J_a=f_{abc}\psi^b\psi^c$ together 
with the $2d$ energy-momentum tensor $T_B$ and the $N=1$ supercurrent $T_F$ 
close into a worldsheet superconformal algebra. The possible gaugings are:
$$
 SU(2)^8,~~ SU(5),~~ SO(7)\times SU(2),~~ G_2\times Sp(4),~~ SU(4)\times SU(2)^3,~~ SU(3)^3.
$$ 
In the following sections we restrict attention to the gauging of maximal rank $H=SU(2)^8_{k=2}$, 
as this is related to the maximal space of deformations.

Respecting the $H_{L}\times H_R=SU(2)^8\times SU(2)^8$ symmetry in Type II 
or the $H_L\times H_R=SU(2)^8\times SO(48)$ in the Heterotic, 
leads to very special tachyon-free constructions with left-right holomorphic 
factorization of the partition function\cite{massivesusy}. 
In the ``prototype'' maximally symmetric constructions all left-moving 
fermions are assigned the same boundary conditions.
The modular invariant (mass generating) partition functions 
in Type II and Heterotic theories are given by
$$
  	Z_{{\rm II}} =\frac{1}{2^2}\sum\limits_{a,b=0,1}{(-)^{a+b}
\frac{\theta[^a_b]^{12}}{\eta^{12}}}\sum\limits_{\bar{a},\bar{b}=0,1}
{(-)^{\bar{a}+\bar{b}}\frac{\bar\theta[^{\bar{a}}_{\bar{b}}]^{12}}{\bar\eta^{12}}}
=(V_{24}-S_{24})(\bar{V}_{24}-\bar{S}_{24})=576,
 $$
 \be
 	Z_{{\rm Het}} =\frac{1}{2}\sum\limits_{a,b=0,1}{(-)^{a+b}
\frac{\theta[^a_b]^{12}}{\eta^{12}}}~\Gamma[H_R] =24\times\left(d[H_R]+[\bar{j}(\bar{z})-744]\right)~.
 \ee
In the Heterotic case there are various possible choices for $H_R$, 
corresponding to $d[H_R] = 1128$ for $H_R=SO(48)$ 
and $d[H_R]=744$ for $H_R=SO(32)\times E_8$ or $H_R=E_8^3$. 
The Klein invariant combination $(\bar j(\bar z)-744)$ is eliminated 
after integration over the fundamental domain (or by imposing level-matching). 

Both $ Z_{\rm II}$ and $ Z_{\rm Het}$ are seen to exhibit a remarkable property: 
the numbers of bosonic and fermionic states are equal at each massive level $n_B-n_F=0$, 
with the exception of the massless modes which are unpaired\footnote{In fact, in the maximally 
symmetric \emph{MSDS} models, the massless spectrum is entirely composed of $576$ scalar bosons 
so that no matching between massless bosonic and fermionic states is possible.} 
(Massive Spectrum Degeneracy Symmetry). 
One may further show that the partition functions of \emph{MSDS} models take the generic form:
\be
Z=m+n(\bar{j}-744),
\ee
where $m,n\in\mathbb{Z}$, and $m=n_B-n_F$ counts the massless level degeneracy. 
The integer $n$ gives the degeneracy of the simple-pole contribution 
from unphysical tachyons in the right-moving sector. 

The \emph{MSDS} structure has a simple CFT interpretation. 
Indeed, it is possible to construct a \emph{chiral} current that 
ensures the mapping of \emph{massive bosonic to massive fermionic} 
representations, while \emph{leaving the massless spectrum invariant}:
\begin{align}
	j_{\alpha}(z) = e^{\frac{1}{2}\Phi-\frac{i}{2}H_0}~C_{24,\alpha}(z),
\end{align}
where $C_{24,\alpha}$ is the spin-field of $SO(24)$ with positive chirality. 
This acquires $(1,0)$ conformal weight once the ghost dressing $-5/8+1/8$ is taken into account. 
Its zero-mode $Q_{MSDS}=\oint{\frac{dz}{2\pi i}~j_{\alpha}(z)}$ defines a 
conserved \emph{MSDS} charge ensuring the mapping of the massive towers of states level by level. 
Acting with the \emph{MSDS} current on the vectorial $\textbf{V}=e^{-\Phi}\hat{\psi}$ 
and the spinorial $\textbf{S}=e^{-\frac{1}{2}\Phi-\frac{i}{2}H_0}S_{24,\alpha}$ representations, 
we find that the \emph{MSDS} current annihilates the massless 
vectorial states $j(z)\textbf{V}(w)=\textrm{regular}$, 
whereas the first massive descendants of the vectorial family $[\textbf{V}]_{(1)}=e^{-\Phi}\mathcal{D}\hat\psi$ 
transform into the spinorial $\textbf{S}$ representation and vice-versa. 
In this notation, $\hat{\psi}\equiv \psi^a\gamma^a$ denotes the contraction 
with the $\gamma$-matrices of $SO(24)$ 
and $\mathcal{D}\equiv\partial+\hat{J}$ is the covariant-like derivative, 
associated to the gauging of supersymmetry and $\hat{J}\equiv\hat\psi\hat\psi$. 
Since only the 24 (chiral) massless states $\psi^a$ are unpaired, 
they contribute to the character difference $V_{24}-S_{24}=24$.

Despite their attractive uniqueness, the maximally symmetric \emph{MSDS} vacua 
are not directly useful for phenomenology. 
As discussed in \cite{massivesusy, reducedMSDS}, 
a direct decompactification of all 8 compact coordinates leads to the conventional 
ten-dimensional superstrings with $\mathcal{N}=8$ supersymmetries ($\mathcal{N}=4$ in the Heterotic case), 
which have no hope of describing chiral matter. 
It was shown in \cite{reducedMSDS} that it is possible to break the 
$H_L\times H_R$ symmetry by considering orbifolds 
of the maximally symmetric vacua that preserve the \emph{MSDS} structure. 
A complete classification of $\mathbb{Z}_2^N$-constructions was given in \cite{reducedMSDS}, 
where the necessary and sufficient conditions for \emph{MSDS} symmetry were derived. 
The result is that whenever the boundary conditions and Generalized GSO-projections (henceforth referred to as GGSO) respect 
the \emph{global, chiral} definition of the spectral-flow operator $j_{{MSDS}}$ 
and if, in addition to the usual modular invariance conditions, the boundary condition vectors $b_a$:
$$
	\psi^a\rightarrow -e^{i\pi b_a}\psi^a,~~\textrm{with}~~b_a\in\{0,1\}
$$
satisfy the extra \emph{holomorphic} constraint:
$$ 
	n_L(b_a)=0(\textrm{mod}~8)~~\textrm{for all}~~b_a,
$$
the resulting vacuum possesses \emph{MSDS} structure. 
It is possible to generalize this construction for more general $\mathbb{Z}_k$-orbifolds, 
but this will be considered elsewhere.

Finally, we briefly comment on the moduli space of the maximally symmetric models. 
The massless spectrum of these models contains $d_L\times d_R$ massless scalars parametrizing the manifold:
\be \label{manifK}
 {\rm\cal K}={SO(d_L,~d_R)\over SO(d_L)\times SO(d_R)}\, ,
\ee
where $d_L,~d_R$ are the dimensions of the $H_L,~H_R$ gauge groups, respectively. 
Not all of these scalars correspond to moduli fields, 
since the symmetry enhancement at the \emph{MSDS} point 
introduces extra massless states into the theory. 
The marginal deformations (flat directions) are those associated 
to the Cartan sub-algebra $U(1)^{r_L}\times U(1)^{r_R}$, 
with $r_L$ and $r_R$ being the ranks of  $H_L$ and  $H_R$, respectively.
As discussed in \cite{massivesusy}, the moduli space of these current-current 
type deformations (modulo dualities) is given by the coset:
\be \label{manifM}
 {\rm\cal M}={SO(r_L,~r_R)\over SO(r_L) \times SO(r_R)}~.
\ee

%%%%%%%%%%%%%%%%%%%%%%%%%%%%%%%%%%%%%%%%%%%%%%%%%%%%%%%%%%%%%%%%%%%%%% 

\section{Marginal deformations of the \emph{MSDS} vacua}\label{marginaldeformations}

In this section we proceed to study marginal deformations of the \emph{MSDS} vacua \cite{massivesusy, reducedMSDS}.
To this extent, we locate these vacua in the moduli space of Type II and Heterotic 
(a-)symmetric orbifold
compactifications \cite{orbifolds} to two dimensions, where supersymmetry is spontaneously broken by geometrical fluxes.
The \emph{MSDS} points are special in that they exhibit enhanced gauge symmetry and massive boson/fermion degeneracy
symmetry. As we will
see, there are lines in moduli space connecting the two-dimensional \emph{MSDS} vacua with four-dimensional
supersymmetric string vacua. Our goal is to analyse the stability of the 
models under marginal deformations along and around these lines,
as well as to exhibit various thermal interpretations. The analysis we perform in this
work is at the
string tree-level (and weak string coupling). 
The more difficult problem of obtaining the backreaction of the quantum corrections due to the
spontaneous breaking of supersymmetry will be left for future investigation.   

\subsection{Maximally symmetric \emph{MSDS} vacuum as a half-shifted lattice}\label{88Lattice}
We begin by considering the Euclidean version of the maximally symmetric \emph{MSDS} vacua. 
In order to obtain the
relevant half-shifted  $\Gamma_{(8,8)}$ lattices \cite{ShifftedLat}, we express 
the partition function of the Type II and Heterotic models as follows:
$$
Z= {V_2\over (2\pi)^2}~\int_\F {d^2\tau\over
4({\rm Im}\tau)^{2}}~ Z_{\rm II,\,Het}~,
$$
\be
Z_{\rm II,\,Het} = \frac{1}{2^2} \sum\limits_{a,b=0,1}{}\sum\limits_{\bar{a},\bar{b}=0,1}{(-)^{a+b}\,
\frac{\theta[^a_b]^4}{\eta^{12}}}\,\Gamma_{(8,8)}[^{a,\bar{a}}_{b,\bar{b}}]\,
{\frac{\bar\theta[^{\bar{a}}_{\bar{b}}]^{4}}{\bar\eta^{12}}}(-)^{(\bar{a}+\bar{b})x}
\left(\frac{ \bar\theta[^{\bar{a}}_{\bar{b}}]^{12} }{ \bar{\eta}^{12}} \right)^{1-x}~,
\ee
where the values $x=1$, $x=0$ correspond to the Type II and Heterotic cases respectively.
Here $V_2$ is the volume associated to the two very large, ``spectator'' directions,
which do not couple to any of the $R$-symmetry charges. 
At the \emph{MSDS} point, the asymmetrically half-shifted $\Gamma_{(8,8)}$ lattice admits 
holomorphic/anti-holomorphic factorization in terms of 
$\theta$-functions:
\be
	\Gamma_{(8,8)}[^{\,a\,,\,\bar{a}\,}_{\,b\,,\,\bar{b}\,}] = \theta[^a_b]^8 ~\bar\theta[^{\bar{a}}_{\bar{b}}]^8~.
\ee 
The Type II and Heterotic moduli spaces are given by 
\be
	\mathcal{M}=\frac{SO\left(8,8+16(1-x)\right)}{SO\left(8\right)\times SO\left(8+16(1-x)\right)},
\ee
modulo the corresponding discrete duality groups. 
To facilitate the analysis of the marginal deformations, we express the lattice in the
familiar Lagrangian form so 
that the values of the $G_{\mu\nu}, B_{\mu\nu}$ moduli at the \emph{MSDS} point can be extracted. 

It is convenient to define $h\equiv a-\bar{a}$ and $g\equiv b-\bar{b}$ and write the lattice as
\be\label{latt1}
	\Gamma_{(8,8)}[^{a,\bar{a}}_{b,\bar{b}}] = \theta[^a_b]^8 ~\bar\theta[^{a-h}_{b-g}]^8.
\ee
Then by using the identity \cite{RK2}
\be
	\theta[^a_b] ~\bar\theta[^{\bar{a}}_{\bar{b}}] =\frac{1}{\sqrt{2\tau_2}} 
\sum\limits_{m,n\in\Z}{e^{-\frac{\pi}{2\tau_2}\left|m+\frac{g}{2}+\tau\left(n+\frac{h}{2}\right)\right|^2+
i\pi\left[mn-\left(a-\frac{h}{2}\right)m+\left(b-\frac{g}{2}\right)n+\frac{h}{2}(b-\frac{g}{2})\right]}},
\ee
we can express (\ref{latt1}) in terms of a product of eight $\Gamma_{(1,1)}$ shifted lattices
\be\label{latt2}
\Gamma_{(8,8)}=\frac{1}{(\sqrt{2\tau_2})^8} \sum\limits_{m_i,n_i\in\Z}{e^{-\frac{\pi}{2\tau_2}\sum\limits_i
\left|m_i+\frac{g}{2}+\tau\left(n_i+\frac{h}{2}\right)\right|^2+i\pi\left[\sum\limits_i{m_i n_i}+
\left(a+\frac{h}{2}\right)\sum\limits_i{m_i}+\left(b-\frac{g}{2}\right)\sum\limits_i{n_i}\right]}}~,
\ee
where the $i$-summation is over the 8 internal directions. 
It is clear from this expression that ordinary supersymmetry 
is spontaneously broken by the couplings of the lattice to the $R$-symmetry charges $(a, b)$ 
and $(\bar{a}, \bar{b})$. In the Type II case, 
these correspond to the left- and right- moving spacetime fermion numbers, $F_L$ and $F_R$, 
respectively, while
in the Heterotic case, the first pair corresponds to the spacetime fermion number $F$ 
and the latter to right-moving gauge $R$-charges. 
Despite the breaking of supersymmetry, the model exhibits massive boson/fermion degeneracy symmetry, 
which will generically be broken by marginal deformations away from the extended symmetry point.  

As is well known, supersymmetry can only be recovered at corners of the moduli space, where certain moduli
take infinite values and the $R$-symmetry charges decouple from the lattice. 
We would like to identify the precise combinations of K\"{a}hler moduli that need to be decompactified
in order to recover supersymmetry. For this purpose, we show in Appendix \ref{AppendixLattice}, 
that the $\Gamma_{(8,8)}$ lattice (equation (\ref{latt2})) can be brought to the following Lagrangian form:
\be\label{FinalLatt} 
	\Gamma_{(8,8)}[^{\,a\,,\,\bar{a}\,}_{\,b\,,\,\bar{b}\,}]=\frac{\sqrt{\det{G_{\mu\nu}}}}{(\sqrt{\tau_2})^{8}}~
\sum\limits_{\tilde{m}^\mu,n^\nu\in\Z}{e^{-\frac{\pi}{\tau_2}
(G+B)_{\mu\nu}(\tilde{m}+\tau n)^\mu (\tilde{m}+\bar{\tau} n)^\nu+i\pi\mathcal{T}}}~,
\ee
where, for later convenience, we have denoted the lattice directions by 
Greek indices taking values from $0,1,\ldots 7$.
$G_{\mu\nu}$ and $B_{\mu\nu}$ stand for the metric and parallelized torsion corresponding to the 
deformed $E_8\times E_8$ lattice. Their values are completely fixed by the \emph{MSDS} symmetry point. 
The phase $\mathcal{T}$ can be written in the form:
\be\label{temperatureRep}
\mathcal{T}_{\textrm{th.}} = \left[\tilde{m}^0 (a+\bar{a})+n^0(b+\bar{b})\right] 
+ \left(\tilde{m}^1 n^1+\tilde{m}^1\bar{a}+n^1\bar{b}\right).
\ee
It describes the couplings of the lattice to the $R$-symmetry charges $(a,b)$ and $(\bar a, \bar b)$.
We see that only two of the eight internal cycles couple to them in this representation. 
As a result, by decompactifying these two cycles, we recover a four-dimensional
supersymmetric vacuum.  
In the Type II case, the $X^0$ cycle is ``thermally'' coupled to the total fermion number $F_L+F_R$, 
whereas the $X^1$ direction is ``thermally'' coupled to the right-moving fermion number $F_R$ \cite{RK,RK2}.  
This particular representation will prove very useful for the subsequent discussions concerning 
the marginal deformations and the thermal interpretation of the Type II \emph{MSDS} models. 
We shall henceforth refer to it as the ``thermal'' or ``temperature'' representation. 

Alternatively, we can bring the lattice in a left-right 
symmetric form by a change of basis: 
$(\tilde{m}^1,n^1)\rightarrow(\tilde{m}^1\pm\tilde{m}^0,n^1\pm n^0)$. 
The new phase-coupling is given by
\be\label{symmetric}
\mathcal{T}_{\textrm{sym}.} = \left(\tilde{m}^0 n^0+\tilde{m}^0 a+n^0 b\right) + \left(\tilde{m}^1 n^1+\tilde{m}^1\bar{a}+n^1\bar{b}\right).
\ee
This basis exhibits the two independent couplings to the 
$R$-charges $(a,b)$ and $(\bar a, \bar b)$.
Now it is in the Heterotic case that the $X^0$ cycle is ``thermally'' coupled to the spacetime fermion number, while
the $X^1$ cycle couples to right-moving $R$-gauge charges.
In the Type II case, the couplings are associated to the  
left- ($F_L$) and right- ($F_R$) moving fermion numbers.
This basis will be henceforth referred to as the ``symmetric'' representation. 
The metric and antisymmetric tensor in the symmetric representation 
are given in equations (\ref{metric}) and (\ref{torsion}) of Appendix \ref{AppendixLattice}.

In either form, the original $(8,8)$-lattice is separated into a $(2,2)$ sub-lattice 
coupled to the $R$-symmetry charges and an ``internal'' $(6,6)$ sub-lattice. 
It is interesting to note that the $(2,2)$ sub-lattice in the symmetric representation 
takes the form of the 
thermal $\Gamma_{(2,2)}$ lattice used in \cite{akpt} to construct tachyon-free thermal models in Type II
theories. The difference here is the 
non-trivial mixing with the $(6,6)$ sub-lattice via the off-diagonal 
$G_{\mu\nu}$, $B_{\mu\nu}$ elements, 
as dictated by the \emph{MSDS} point in moduli space. This ``decomposed'' form is in fact 
the most general possible form of an $(8,8)$ lattice ``thermally'' coupled to the left- and right- moving 
fermion numbers $F_L$, $F_R$. It is the starting point for the discussions that follow.

%%%%%%%%%%%%%%%%%%%%%%%%%%%%%%%%%%%%%%%%%%%%%%%%

\subsection{The creation of spacetime and the \emph{MSDS}/4d gauged supergravity correspondence}

The physical interpretation of the Euclidean model depends on the
choice of the time direction. We first consider
two-dimensional cold \emph{MSDS} vacua, where the Euclidean
time is taken along one of the non-compact, spectator
directions. Upon rotation to Lorentzian signature, these
two directions analytically continue to the longituninal lightcone directions $X^{\pm}$.
The Euclidean partition function determines the energy density of the vacuum,
which is induced at the one-loop (and higher genus) level by
the spontaneous breaking of supersymmetry.  

In the previous section, the $\Gamma_{(8,8)}$  
lattice corresponding to the maximally symmetric \emph{MSDS} vacuum 
was expressed in its fully ``decomposed'' form (\ref{temperatureRep}), 
where only a $(2,2)$ sub-lattice couples 
to the $R$-symmetry charges $(a+\bar a,b+ \bar b)$ and $(\bar{a}, \bar{b})$. 
In that notation, the $(2,2)$ sub-lattice
is spanned by the quantum numbers $[(\tilde{m}^0,n^0);\,(\tilde{m}^1,n^1)]$. 
We denote the associated compact directions by $X^0$ and $X^1$
respectively. In Type II theories, the $R$-symmetry charges correspond to the total and 
right-moving spacetime fermion numbers,
$F_L+F_R$ and $F_R$. Their coupling to the lattice 
breaks the initial $(4,4)$ supersymmetries to $(0,0)$ spontaneously. 
In the infinite-radius limit, $R_0,\, R_1\to \infty$, where the $X^0$ and $X^1$ cycles decompactify, only the
$(\tilde{m}^0,n^0)=(0,0)$ and $(\tilde{m}^1,n^1)=(0,0)$ orbits survive, 
and the $R$-symmetry charges effectively decouple from the
lattice. In this limit we recover a four-dimensional maximally supersymmetric, ${\cal N}=8$, Type II vacuum. 
As we will show in Section \ref{stability}, this particular decompactification limit can be described 
in terms of adiabadic motion along a \emph{tachyon-free trajectory} in moduli space, 
which interpolates continuously between the two-dimensional \emph{MSDS} vacua 
and the four-dimensional supersymmetric ones.

The decompactification limit can be achieved via large marginal 
deformations of the current-current type. Specifically, one perturbs the sigma-model action by adding marginal operators of the form
\be\label{deform1}
	\lambda_0 ~J_0\times\bar{J}_0+\lambda_1 ~J_1\times\bar{J}_1\sim \lambda_0~\partial X^{0}\bar\partial X^{0}
+\lambda_1~\partial X^{1}\bar\partial X^{1}~,
\ee 
and then takes the $\lambda_0,\lambda_1 \to \infty$ limit adiabatically. Of course, we would like to 
identify more trajectories 
along which we do not encounter tachyonic instabilities as we deform from the non-singular \emph{MSDS}
vacuum to the supersymmetric one. This problem will be investigated in Section \ref{stability}.  

Near the extended symmetry point, the corresponding
$T^2$ sub-manifold is threaded by non-trivial ``gravito-magnetic'' fluxes, 
as indicated by the non-vanishing values of the 
$G_{0I}, G_{1I}$ and
$B_{0I}, B_{1I}$, $I=1\dots 6$, moduli and the couplings to the
$R$-symmetry charges. When the size of space is of the order of the
string scale, classical  
notions such as geometry, topology and even the dimensionality of the 
underlying manifold are ill-defined (See \cite{CosmoTopologyChange} for
discussions). 
A familiar example, the $SU(2)_{k=1}$ WZW
model, serves to illustrate these issues: the target space can be taken to be a highly curved 
three-dimensional sphere with one unit of NS-NS 3-form flux, 
or in an equivalent description, a circle at the self-dual radius. For large level $k$, however,
the target space can be unambiguously described as a large 3-sphere with NS-NS 3-form flux.  
Similarly, in our case, a clear geometrical picture involving four large spacetime dimensions, 
along with an effective field theory description,
arises in the large moduli limit.
Two additional large dimensions {\it emerge} via marginal deformations of the current-current type.
In a cosmological setting, the deformation moduli acquire time-dependence.
It is plausible then that a four-dimensional cosmological space is created dynamically 
from an initially non-singular, $2d$ \emph{MSDS} vacuum.  

The four-dimensional maximally supersymmetric vacuum (obtained in the infinite-radius limit) corresponds to a
Type II compactification on a $T^6$ torus with fluxes, with the $T^6$ moduli remaining close to the fermionic point. 
The internal moduli can be also deformed, and so one obtains a rich manifold of maximally supersymmetric vacua. 
The low energy effective description is in terms of $d=4$, ${\cal N}=8$ gauged supergravity, 
where the gauging
is induced by the internal fluxes. This illustrates the correspondence of the maximally symmetric \emph{MSDS} vacuum 
with the maximal $d=4$, ${\cal N}=8$ gauged supergravity. This correspondence can be also extended to less symmetric cases, 
where the
initial \emph{MSDS} symmetry is reduced by orbifolds \cite{reducedMSDS}:
\be
 \left[{\rm Z_{\rm orb} }: ~MSDS~\right]_{d=2 } ~~\longleftrightarrow~~  \left[~{\cal N}\le 8 : ~SUGRA~\right]_{d=4,\, {\rm fluxes}}~.
\ee
This more general class of models includes also four-dimensional $(4,0)$ and $(0,4)$ vacua, as well as their orbifolds, 
obtained by decompactifying a suitable cycle within the $(2,2)$ sub-lattice 
together with one of the directions associated with the $(6,6)$ sub-lattice. 

In the Heterotic models, only the left-moving spin structures $(a,b)$ are associated with the spacetime
fermion number and, 
thus, only one modulus need be infinitely deformed in order to yield a supersymmetric theory.
Starting with the maximally symmetric \emph{MSDS} Heterotic model, 
a four-dimensional ${\cal N}=4$ model can be obtained by decompactifying one additional direction.
A novel feature is the coupling of the $X^1$ cycle to internal, right-moving gauge charges. This coupling is induced
by discrete Wilson lines associated with the non-abelian gauge field (arising from the right-moving
sector). It would be interesting to identify initially \emph{MSDS} Heterotic vacua, which can be connected via large
marginal deformations to phenomenologically viable, $d=4$, ${\cal N}=1$ chiral models. The classification of
the ${\cal N}=1$ models and their phenomenology will be studied elsewhere \cite{chiralhet}.

At intermediate values of the radii $R_0,R_1$, the four-dimensional spacetime part includes 
non-trivial geometrical fluxes which induce the spontaneous breaking of supersymmetry.
In the maximally symmetric Type II models there are two supersymmetry breaking scales associated with the breaking
of the left- and right- supersymmetries respectively. 
In a cosmological setting, one should also introduce temperature into the system.
The conventional thermal deformation is obtained as follows. 
One starts with a cold \emph{MSDS} vacuum where the Euclidean time is along 
one of the two non-compact spectator directions, and compactifies it on a circle. 
This circle must be coupled to the total fermion number $F$ in a manner consistent
with the spin/statistics connection. The induced temperature     
is inversely proportional to its radius, and
leads to the familar Hagedorn instabilities when it becomes of the order of the string scale.
As we will explicitly show in the next
section, however, some Euclidean versions of the models, where the two
spectator dimensions remain non-compact, admit a thermal interpretation of their own. 
They describe thermal string vacua 
in the presence
of left/right asymmetric gravito-magnetic fluxes, very similar 
in fact to the tachyon-free thermal models studied in \cite{akpt}.
In these Euclidean models one of (the already existing) supersymmetry breaking scales 
is identified with the temperature.

%%%%%%%%%%%%%%%%%%%%%%%%%%%%%%%%%%%%%%%%%%%%%%%%%

\subsection{Thermal interpretation of the models}\label{thermal}
 
In this section, 
we consider Euclidean versions of Type II \emph{MSDS} models, 
where time is identified 
with one of the compact directions associated with the $T^8$ torus.
The physical interpretation of such a model is  
in terms of a string thermal ensemble, with the temperature 
given by the inverse radius of the Euclidean time cycle.
Consistency with the spin/statistics connection and modular invariance require
that the quantum numbers associated with the Euclidean time cycle couple to the
spacetime fermion number $F$ through a specific cocycle. 
At the level of the one-loop thermal
amplitude, this cocycle takes the form
\be
e^{i\pi(\tilde m^0 F+ n^0 \tilde F)}~,
\ee 
correlating the string winding numbers with the spacetime spin \cite{RK, RK2}.

In the \emph{MSDS} models, we can identify a thermal cycle, which couples to the spacetime fermion
number as dictated by the spin/statistics connection, within the $(2,2)$ sub-lattice responsible for the supersymmetry
breaking. However, as in the models of \cite{akpt}, this cycle is also threaded by 
non-trivial ``gravito-magnetic'' fluxes, associated
with various $U(1)$ graviphoton and axial vector gauge fields. The presence of these fluxes {\it refines}
the thermal ensemble, and so chemical potentials associated with the fluxes appear \cite{akpt}.
Despite the fact that the radius of the Euclidean time circle is smaller than the Hagedorn one,
these additional fluxes render the partition function at the \emph{MSDS} points finite. 
In light of the thermal interpretation of the models, we motivate them further.   

At finite temperature, the one-loop string partition function diverges when
the radius of the Euclidean time circle is smaller than the Hagedorn value: $R_0 < R_H$.
This divergence can be associated with the exponential
growth of the density of single-particle string states as a function of mass. 
It can also be interpreted
as an infrared divergence since, precisely for $R_0 < R_H$, certain stringy states 
winding around the Euclidean time
circle become tachyonic. 
These instabilities signal the onset of a non-trivial phase transition at around the Hagedorn
temperature \cite{AW,AKADK}. If a stable high temperature phase exists, it should be 
reachable through the process of
tachyon condensation, where the tachyon ``rolls'' to a new stable vacuum. In the literature there
are many proposals concerning the nature of this phase transition, but 
the large values of the tachyon condensates involved drive
the system outside the perturbative domain, and so adequate quantitative
description of the dynamics is lacking (see eg \cite{AW,AKADK,BR}). 

Another way to proceed is to deform the background via other types
of condensates that can be implemented at the string perturbative level,
so that the tachyonic instabilities are removed.
In addition to the temperature
deformation, we turn on certain discrete gravito-magnetic fluxes threading the Euclidean time circle,
as well as other internal spatial circles \cite{akpt}. 
In fact, these fluxes can be described in terms of gauge field condensates,
of zero field strength, but with non-zero value of the Wilson line around the
Euclidean time circle.
The would-be winding tachyons get lifted 
as they are charged under the corresponding gauge fields. 
An analogue to keep in mind is the case of a charged particle moving on a plane, under
the influence of an inverted harmonic oscillator potential. A large enough magnetic field can stabilize the motion
of the particle. The thermal \emph{MSDS} vacua correspond precisely to such tachyon-free setups. 
The \emph{MSDS} algebra completely fixes the structure of the corresponding statistical ensemble,
determining the values of the temperature and the chemical potentials
associated with the gravito-magnetic fluxes.

The thermal vacua are of particular interest in the context of string cosmology. 
It was argued in \cite{massivesusy}, that the \emph{MSDS} vacua are promising candidates to 
describe the very early phase of the Universe. As the thermal \emph{MSDS} vacua are free
of Hagedorn instabilities, they could be the starting point of the cosmological
evolution. In such a dynamical setting, the physical moduli acquire time dependence, so it is
necessary to construct thermal \emph{MSDS} vacua that remain tachyon-free under 
fluctuations of the dynamical moduli, as well as to identify safe trajectories in moduli
space that connect the models to higher-dimensional thermal ones. 

Now not all of the $64$ moduli in the $\frac{SO(8,8)}{SO(8)\times SO(8)}$ space 
correspond to fluctuating fields. Local worldsheet symmetry permits one to gauge away
the oscillator degrees of freedom associated to the Euclidean time direction
and to an additional spatial direction. This implies in
particular that certain combinations involving the $G_{0I}$ and $B_{0I}$ moduli can be frozen,
and the corresponding fluctuations are set to zero. 
As we will see, the various geometrical data associated to the Euclidean time direction 
are naturally interpreted as the thermodynamical parameters of the statistical ensemble. 
For the maximally symmetric \emph{MSDS} models, we choose the direction of Euclidean time to 
lie in the first direction, $X^0$, of the $(2,2)$ sub-lattice, see equation (\ref{symmetric}), 
coupling to the left-moving $R$-symmetry charges. 
As before, the coordinates of the full $(8,8)$-lattice are collectively 
denoted by Greek characters $\{\mu=0,1,\ldots,7\}$; the coordinates $X^I$, other than 
the Euclidean time will be denoted by Latin characters $\{I=1,2,3,\ldots,7\}$. 
This includes, in particular, the direction $I=1$ in the $(2,2)$ sub-lattice 
coupling to the right-moving $R$-symmetry charges.

%%%%%%%%%%%%%%%%%%%%%%%%%%%%%%%%%%%%%%%%%%%%%%%%%%%%%%%%%%%%%%%%%%%%%%%%%%%%%%%%%%%%%
%%%%%%%%%%%%%%%%%%%%%%%%%%%%%%%%%%%%%%%%%%%%%%%%%

\subsubsection{The thermal ensembles}\label{ensembles}

In this section we exhibit the thermal interpretation of the class of models considered in this paper. 
Our discussion will be focused on the maximally symmetric \emph{MSDS} models, 
but may be suitably adapted to apply to the twisted \emph{MSDS} and Hybrid models of the following sections. 
Our starting point will be a general lattice of the form (\ref{NewLattice}):
\be
	\Gamma_{(8,8)}[^{\,a\,,\,\bar{a}\,}_{\,b\,,\,\bar{b}\,}]=\frac{\sqrt{\det{G_{\mu\nu}}}}{(\sqrt{\tau_2})^{8}}~
\sum\limits_{\tilde{m}^\mu,n^\nu\in\Z}{e^{-\frac{\pi}{\tau_2}
(G+B)_{\mu\nu}(\tilde{m}+\tau n)^\mu (\tilde{m}+\bar{\tau} n)^\nu+i\pi\mathcal{T}}}.
\ee
We use the tilde and upper indices ($\tilde{m}^\mu, n^\mu$) to 
denote the winding numbers around the two non-trivial loops of the worldsheet torus.
The momentum quantum numbers ($m_{\mu}$) are denoted without 
tildes and with lower indices. 

We single-out the direction $(\tilde{m}^0,n^0)$ of the lattice, 
corresponding to the compact Euclidean time direction. 
We assume the following general form for the metric $G_{\mu\nu}$ and antisymmetric tensor $B_{\mu\nu}$:
\be\label{metricDecomposition}
	G_{\mu\nu}=
\left(\begin{array}{c|c}
	G_{00} & G_{0J} \\ \hline
	G_{I0} & G_{IJ} \\
\end{array}\right)~~,~~B_{\mu\nu}=
\left(\begin{array}{c|c}
	0 & B_{0J} \\ \hline
	B_{I0} & B_{IJ} \\
\end{array}\right).
\ee
For the case of the maximally symmetric \emph{MSDS} models, 
their numerical values are given in (\ref{metric}) and (\ref{torsion}).
The phase coupling $\mathcal{T}$ takes the form
\be
	\mathcal{T}= (\tilde{m}^0 n^0 + a \tilde{m}^0 +b n^0) + (\tilde{m}^1 n^1+ \bar{a}\tilde{m}^1 + \bar{b} n^1).
\ee
For later convenience, the determinant may be written as
\be\label{determinant}
	\det{G_{\mu\nu}} = R_0^2~\det{G_{IJ}},
\ee
where $R_0^2$ is given by\footnote{Throughout, $G^{IJ}$ is the inverse of $G_{JK}$: $G^{IJ}G_{JK}=\delta_K^I$.}
\be
	R_0^2 ~\equiv~ G_{00}-G_{0I}G^{IJ}G_{J0},
\ee
and will be shown below to correspond precisely to the inverse temperature $\beta=2\pi R_0$. 

In this form the $X^0$ and $X^1$ directions couple (independently) to left- ($F_L$) and right- ($F_R$) 
moving fermion numbers respectively. 
To make the thermal nature of the coupling explicit, 
we change basis by shifting $(\tilde{m}^1,n^1)\rightarrow (\tilde{m}^1-\tilde{m}^0, n^1-n^0)$, 
so that the $X^0$-cycle couples to the total spacetime fermion number $F_L+F_R$:
\be\label{coupling3}
	\mathcal{T}\rightarrow [(a+\bar{a})\tilde{m}^0+(b+\bar{b})n^0]
+[\tilde{m}^1 n^1+\bar{a} \tilde{m}^1+\bar{b}n^1]+(\tilde{m}^0 n^1-\tilde{m}^1 n^0)
=\mathcal{T'}+(\tilde{m}^0 n^1-\tilde{m}^1 n^0).
\ee
The new phase coupling $\mathcal{T}'$ is the sum of a thermal coupling to the total fermion number $F_L+F_R$, 
which corresponds precisely to the conventional thermal deformation of Type II theories, 
along with an independent coupling of the ${X'}^1$ direction to $F_R$. 
The latter deformation breaks spontaneously the right-moving supersymmetries, $\mathcal{N}=(4,4)\rightarrow(4,0)$, 
before finite temperature is introduced into the theory. 
The thermal deformation breaks the remaining left-moving supersymmetries to $(0,0)$.
The last term in the transformation (\ref{coupling3})  
is the discrete torsion contribution to $B'_{01}$.
In the new basis, the metric and antisymmetric tensor are given by
\be
		G_{\mu\nu}'=
\left(\begin{array}{c|c}
	G_{00}+2G_{01}+G_{11} & G_{0J}+G_{1J} \\ \hline
	G_{I0}+G_{I1} & G_{IJ} \\
\end{array}\right)~~,~~B_{\mu\nu}'=
\left(\begin{array}{c|c}
	0 & B_{0J}+B_{1J}+\frac{1}{2}\delta_{1J} \\ \hline
	B_{I0}+B_{I1}-\frac{1}{2}\delta_{I1} & B_{IJ} \\
\end{array}\right).
\ee
The radius $R_0^2$ is left invariant by the change of basis:
\be
	G_{00}'-G_{0I}'{G'}^{IJ}G_{J0}'=G_{00}-G_{0I}G^{IJ}G_{J0}=R_0^2.
\ee

Next we Poisson re-sum all the $\tilde{m}^I$ windings transverse to the temperature direction. 
This sets the $(7,7)$ lattice associated with $X'^{I}$ directions to its Hamiltonian picture. 
The full $(8,8)$ lattice reads:
$$
	\frac{R_0}{\sqrt{\tau_2}}\sum\limits_{\tilde{m}^0,n^0\in\Z}
{e^{-\frac{\pi R_0^2}{\tau_2}|\tilde{m}^0+\tau n^0|^2}~(-)^{(a+\bar{a})\tilde{m}^0+(b+\bar{b})n^0}}
$$
\be\label{thermallattice}
	\times\sum\limits_{m_I,n^I\in\Z}{q^{\frac{1}{2}P_L^2}\bar{q}^{\frac{1}{2}P_R^2}~
(-)^{\bar{b}n^1}}~e^{2\pi i\tilde{m}^0({G'_{0I}}Q^I_{(M)}-{B'_{0I}}Q^{I}_{(N)})}~.
\ee
Here, $P_{L,R}^2\equiv G_{IJ}P^I_{L,R} P^J_{L,R}$ are the canonical momenta of the transverse lattice:
\be
	P_{L,R}^I = \frac{1}{\sqrt{2}}~G^{IJ}~\left(m_J+\frac{\bar{a}+n^1}{2}\delta_{J1}+(B_{J\mu}'\pm G_{J\mu}')n^\mu\right),
\ee
with $(+)$ and $(-)$ corresponding to the left- and right- moving canonical momenta, respectively. 
We have also defined 
\be 
Q_{(M)}^I \equiv \frac{1}{\sqrt{2}}\left(P_L^I + P_R^I\right),~~~~ Q_{(N)}^I \equiv \frac{1}{\sqrt{2}}\left(P_L^I - P_R^I\right).
\ee

The thermal interpretation of the models becomes manifest, when we decompose
the lattice (\ref{thermallattice}) in modular orbits. These are the $(\tilde m^0,n^0)=(0,0)$
orbit which is integrated over the fundamental domain, and the $(\tilde m^0,0),\,\, \tilde m^0\ne 0$ orbit
which is integrated over the strip\footnote{For sufficiently large $R_0$
to ensure absolute convergence. See also the discussion in Section \ref{thermalHybrid}.} \cite{akpt}. 
At zero temperature, we recover a supersymmetric 3-dimensional $(4,0)$ model, and
so the contribution of the $(0,0)$ orbit vanishes.
The strip integral determines the free energy of the model at one loop.
The integrand involves
$$
	\frac{R_0}{\sqrt{\tau_2}}\sum_{\tilde{m}^0\ne 0}
e^{-{\pi \over \tau_2}(R_0 \tilde{m}^0)^2}~(-)^{(a+\bar{a})\tilde{m}^0}
$$
\be
	\times\sum\limits_{m_I,n^I\in\Z}{q^{\frac{1}{2}P_L^2}\bar{q}^{\frac{1}{2}P_R^2}~
(-)^{\bar{b}n^1}}~e^{2\pi i\tilde{m}^0({G'_{0I}}Q^I_{(M)}-{B'_{0I}}Q^{I}_{(N)})}~,
\ee
where now the $14$ transverse $U(1)$ charges $Q_{(M)}^I,\, Q_{(N)}^I$ (evaluated at $n^0=0$) 
are those associated with the graviphoton, $G'_{I\mu}$, and axial vector, $B'_{I\mu}$, fields.
 
From the above expression, we can infer
the form of the complete spacetime partition function:
\be\label{thermaltrace}
	\mathcal{Z}(\beta,G'_{0I},B'_{0I})=\textrm{tr}{\left[~e^{-\beta H}~
e^{2\pi i\left(G'_{0I} {Q}^{I}_{(M)}-B'_{0I} {Q}^I_{(N)}\right)}\right]},
\ee
where the trace is over the Hilbert space of the 3-dimensional $(4,0)$ 
theory\footnote{Since the Hamiltonian is quadratic in the charges, the
partition function is real.}.
The model describes a thermal ensemble at temperature $T^{-1}=\beta=2\pi R_0$, which is 
further deformed by 
(imaginary) chemical potentials for the graviphoton and axial vector charges
$Q_{(M)}^I,\, Q_{(N)}^I$. At temperatures sufficiently below the string scale, with the other
compact spatial cycles held fixed at the fermionic point, the massive states carrying these charges 
effectively decouple from 
the thermal system, and so in this limit we get a conventional thermal ensemble.

Notice that the argument of the phase in (\ref{thermaltrace}) is scale invariant.
To make this invariance explicit, we rewrite it in terms of the integral
charges 
\be
	\hat{m}_{I} \equiv m_I+{\bar a +n^1 \over 2}\delta_{1I}~,~~~~n^I~.
\ee
We obtain that
\be\label{ThermalInterpretationMAX}
	\mathcal{Z}(\beta,\mu^{I},{\tilde \mu}_I)=\textrm{tr}{\left[~e^{-\beta H}~
e^{2\pi\left(\mu^{I} \hat{m}_{I}-{\tilde \mu}_{I} n^I\right)}\right]}~,
\ee
where $\mu^{I}\equiv i G'_{0K}G^{KI}$, ${\tilde \mu}_I\equiv i (B'_{0I}-G'_{0J}G^{JK}B_{KI})$ are the chemical potentials 
for the charges $\hat m_I,\, n^I$.
The background expectation values 
$$
	\hat{G}_0^I\equiv G'_{0K}G^{KI},
$$
\be\label{ChemPotentialGB}
	\hat{B}_{0I} \equiv B'_{0I}-\hat{G}_0^K B_{KI},
\ee
associated to the Euclidean time direction are non-fluctuating 
thermodynamical parameters (chemical potentials) of the statistical ensemble.

%%%%%%%%%%%%%%%%%%%%%%%%%%%%%%%%%%%%%%%%%%%%%%%%

\subsection{Deformations and stability}\label{stability}

In this section we investigate
whether the \emph{MSDS} vacua are stable under marginal deformations;
that is, whether small (or even large) deformations in moduli space around the \emph{MSDS} point remain 
free of tree-level tachyonic instabilities. The answer depends on the number of deformable moduli, 
their relative size, as well as on the interpretation of the model, which may describe 
a two-dimensional string model on $\mathcal{M}^{2}\times T^8$ or a string thermal ensemble. 

In the remaining part of this section we will be concerned with the stability properties 
of the maximally symmetric \emph{MSDS} vacua under arbitrary marginal deformations 
of the current-current type. 
Our analysis focuses on thermal Type II vacua, but it can be easily extended to the other cases. 
We show that the maximally symmetric \emph{MSDS} models are unstable under 
arbitrary deformations of the dynamical moduli.
Note that for the two-dimensional (cold) maximally symmetric vacua,
the negative curvature induced once we take the backreaction of 
the quantum corrections into account,
may lift such tree-level ``tachyonic instabilities'' (which may occur as we deform
away from the \emph{MSDS} point).   
We also identify the necessary conditions for more general thermal
vacua to be stable under arbitrary deformations of the dynamical moduli. 
We show that only for a very specific thermodynamical phase (choice of chemical potentials) 
are the corresponding thermal vacua stable under arbitrary marginal deformations of the transverse moduli and the temperature. 
This restriction can be waived by considering orbifold twists of the original model, 
which project out some of the deformation moduli. Stable orbifold \emph{MSDS} vacua 
will be constructed in Section \ref{twisted} of the paper. 

Tachyons in the string spectrum may only appear in the $\textrm{NS-NS}$ sector, where $a=\bar a=0$. 
In terms of $SO(8)$-characters, the `dangerous' sector in the partition function 
is the sector with odd chiral and anti-chiral GSO projections to the lattice:
\be\label{dangerousSector}
	O_8\bar{O}_8\, \frac{1}{2}\sum\limits_{b',\bar{b}'=0,1}{(-)^{b'+\bar{b}'}\,
\Gamma_{(8,8)}[^{~0\,,\,0~}_{~b'\,,\,\bar{b}'~}](G_{IJ},B_{IJ})}.
\ee
The odd GSO projection is imposed independently to the left- and right-movers by the $b',\bar{b}'$-summations.
It is clear that only lattice BPS states can become tachyonic, 
since any left-moving (right-moving) oscillator would immediately make the state massless, 
at least holomorphically (anti-holomorphically). At the \emph{MSDS} point, 
where the moduli take the precise values given in equations (\ref{metric}) and (\ref{torsion}), 
the lowest lying physical states in the NS-NS spectrum are all massless. 
The question we would like to address is what happens to these particular states under arbitrary marginal deformations. 

The metric and antisymmetric tensor will be assumed to be in the ``temperature basis", 
where the phase coupling has the form
\be\label{thermalCoupling}
	\mathcal{T} = \left[(a+\bar{a})\tilde{m}^0 + (b+\bar{b})n^0\right] 
+ \left[\tilde{m}^1 n^1+\bar{a}\tilde{m}^1+\bar{b}n^1\right].
\ee
The primes in the metric and antisymmetric tensor used in the previous section 
are here omitted for notational simplicity.
Recall from the previous section that the temperature modulus was identified with $R_0^2=G_{00}-G_{0I}G^{IJ}G_{J0}$. 
We can use vielbein notation that brings the lattice metric to the flat frame $G_{\mu\nu}=e^{a}_{\mu}e^{a}_{\nu}$. 
It is a well-known fact in linear algebra that 
any positive definite symmetric matrix $G_{\mu\nu}$ has a \emph{unique} 
decomposition\footnote{This decomposition is a special 
case of the standard $LU$-decomposition and is called Cholesky decomposition.} 
into the product of an upper and a lower triangular matrix. 
Indeed, let us choose the vielbein matrix $(e^a)_\mu$ to be the unique lower triangular matrix in this decomposition. 
The first row of this vielbein matrix,
\be
	(e^{a=0})_{\mu} = ( R_0, 0,\ldots,0),
\ee	
corresponds precisely to the radius $R_0$ associated to the temperature. 
The inverse metric is similarly decomposable $G^{\mu\nu}=(e^{*a})^\mu(e^{*a})^\nu$ in terms of 
the dual vielbein lattice base vectors $(e^{*a})^\mu$, 
which can be equivalently seen as an upper triangular matrix. The first row of the dual vielbein matrix is given by
\be
	(e^{*,a=0})^\mu = \frac{1}{R_0}\left(1, -{\hat{G}_0}^J\right),
\ee 
with $J=1,\ldots,7$, and 
\be
	{\hat{G}_0}^J \equiv G_{0I} G^{IJ}
\ee
are the (frozen) chemical potentials (\ref{ChemPotentialGB}) 
associated with the $U(1)$ graviphoton charges. 
Similarly, we define the antisymmetric tensor $b^{ab}$ in the flat frame using the decomposition 
$B_{\mu\nu}=e^a_\mu e^b_\nu b^{ab}$. We then find
\be
	{b^0}_K \equiv b^{0b} e^b_K = \frac{1}{R_0}\left( B_{0K}-{\hat{G}_0}^J B_{JK}\right) = \frac{\hat{B}_{0K}}{R_0},
\ee
which were identified in the previous section as the chemical potentials associated with the axial $U(1)$ charges. Similarly:
$$
	{b^0}_0 \equiv b^{0b}(e^b)_0 =(e^{*,0})^I B_{I 0} = \frac{\hat{G}_0^I \hat{B}_{0I}}{R_0}.
$$
From the point of view of the thermal interpretation, 
this particular vielbein decomposition is natural because it 
directly reveals the temperature and chemical potentials of the statistical system.

In the flat vielbein frame, the mass formula for the BPS-spectrum becomes
\be\label{massFormula}
	\frac{1}{2}P_{L,R}^2 = \frac{1}{4}\sum\limits_{a=0}^{7}
{\left( (e^{*a})^\mu(\hat{m}_\mu+B_{\mu\nu}n^\nu) \pm e^a_\mu n^\mu\right)^2},
\ee
where $\hat{m}_\mu \equiv m_{\mu} +\frac{1}{2}(\bar{a}+n^1)\delta_{\mu,1}$. 
The half-shifted momentum, $m_1$, corresponds to the direction 
associated with the breaking of the right-moving supersymmetries.  
The decomposition of the mass formula into 8 perfect squares is very useful for the subsequent stability analysis.
We first examine the contribution of the first square in (\ref{massFormula}), involving the temperature radius $R_0$:
\be\label{firstSquare}
	\frac{1}{4}\left(\frac{m_0-{\hat{G}_0}^I \hat{m}_I
+\hat{B}_{0J}n^J+\hat{B}_{0I}\hat{G}^I_0 n^0}{R_0} \pm R_0 n^0\right)^2. 
\ee

We note that contribution (\ref{firstSquare}) is entirely determined by the thermodynamical 
variables of the statistical ensemble and is 
unaffected by the fluctuations of the dynamical moduli. 
The fluctuations of the latter will only affect the 7 remaining contributions in (\ref{massFormula}), 
organized into squares associated to the ``transverse'' moduli space. 
Now arbitrary fluctuations of the dynamical moduli 
can always lower such a ``transverse'' contribution to its minimum (vanishing) value. 
Consequently, if the model under consideration is to be free of tachyonic 
instabilities for \emph{any} deformations 
of the dynamical moduli, 
the frozen contribution (\ref{firstSquare}) 
should suffice to render the low-lying spectrum at least \emph{chirally massless}
\footnote{By chirally massless we mean states that are massless either from the holomorphic 
or anti-holomorphic sector of the theory. This is because only level-matched states are physical.}. 
Next we examine under what conditions this is the case.

Imposing independent odd left- and right-moving GSO projections as in (\ref{dangerousSector}) yields:
\be
	n^0 \in 2\Z+1~~\textrm{and}~~	n^1 \in 2\Z.
\ee
The level-matching condition is then written as
\be
	m_0 n^0 + \left(m_1+\frac{n^1}{2} \right)\,n^1 + m_2 n^2+\ldots + m_7 n^7 =0.
\ee
The thermal contribution (\ref{firstSquare}) to the mass of a state with charges $(m_\mu,n^\mu)$ has a minimum at radius
\be
	R_{0,\min} = \left| \frac{m_0-{\hat{G}_0}^I \hat{m}_I+\hat{B}_{0J}n^J+\hat{B}_{0I}\hat{G}^I_0 n^0}{n^0}\right|,
\ee
and the minimum, yet non-vanishing, value of $\frac{1}{2}\,P_{L,R}^2$ is given by
\be
	\frac{1}{2}P_{\min}^2 =\left|\left(m_0-{\hat{G}_0}^I \hat{m}_I
+\hat{B}_{0J}n^J+\hat{B}_{0I}\hat{G}^I_0 n^0\right)\, n^0\right| .
\ee
A thermal model that is tachyon-free regardless of deformations along 
the ``transverse" moduli space should, necessarily, have $\frac{1}{2}\,P_{\min}^2\geq \frac{1}{2}$. 
It is straightforward to show, by considering various states $(m_\mu,n^\mu)$ that satisfy level-matching, 
that the corresponding conditions on the chemical potentials are:
$$
\left.\begin{array}{c}
\hat{G}_{0}^k \in \mathbb{Z} \\
\hat{B}_{0k} \in \mathbb{Z} 
\end{array}\right\}~~,~~\textrm{for}~k=2,\ldots 7,
$$
\be\label{TachyonFreeCond}
\hat{G}_{0}^1\in 2\mathbb{Z}+1~~,~~ \hat{B}_{01} \in \mathbb{Z}+\frac{1}{2} ~.
\ee
As a result, any Type II vacuum 
in a thermodynamical phase satisfying (\ref{TachyonFreeCond}) 
will be tachyon-free for any value of the temperature $R_0$ and 
for all expectation values of the ``transverse'', dynamical moduli. 
It is shown in Appendix \ref{conditionsCheck} that, whenever these conditions are met, the chemical potentials can always be
rotated  (by a discrete $O(8,\mathbb{Z})\times O(8,\mathbb{Z})$ rotation) to the form $\hat{G}_{0}^{k} =
\hat{B}_{0k} = 0$, $\hat{G}_{0}^{1} = 2\hat{B}_{01} = \pm 1$.
This is
precisely the case when the temperature cycle $S^1$ can be factorized from the
remaining toroidal manifold, in which case it couples to the
left-moving fermion number $F_L$ only, as in the models of \cite{akpt}.
It is straightforward to show that the spacetime partition function (\ref{ThermalInterpretationMAX})
reduces in this case to
\be\label{KounnasIndex}
	\mathcal{Z}=\textrm{tr}{\left[~e^{-\beta H}~
(-1)^{F_R}\right]}~,
\ee
the right-moving fermion index.

For the maximally symmetric \emph{MSDS} models, the thermodynamical parameters 
are numerically determined to be:
$$
	R_0 = \frac{1}{4},
$$
$$
	\hat{G}_0^I=\left(\frac{3}{2},\frac{1}{4},0,\frac{1}{4},\frac{1}{2},\frac{1}{4},0\right),
$$
\be
	\hat{B}_{0I}=\left(0,-\frac{3}{4},-\frac{1}{4},-\frac{3}{4},0,-\frac{1}{4},0\right).
\ee
They obviously do not satisfy conditions (\ref{TachyonFreeCond}) and so, 
the maximally symmetric \emph{MSDS} models are not tachyon-free for arbitrary marginal deformations,
including small marginal deformations around the non-singular point. 

However, we can now show that the asymptotic limit $R_0\rightarrow\infty$ 
continuously interpolates between the thermal \emph{MSDS} vacuum and supersymmetric vacua 
at zero temperature, without encountering tachyonic states. For this purpose we notice 
that the (shifted) momentum contribution 
$\hat{M}_0\equiv m_0-{\hat{G}_0}^I \hat{m}_I+\hat{B}_{0J}n^J+\hat{B}_{0I}\hat{G}^I_0 n^0 $ comes as
\be
	|\hat{M}_0|=\left|\Z+\frac{\Z}{2}
+\frac{\Z}{4}\right|\in\left\{0,\frac{1}{4},\frac{1}{2},\ldots\right\}.
\ee
Since $n^0$ is odd, only states with $n^0=\pm 1$ winding number can become tachyonic. 
Clearly, only the two cases $\hat{M}_0=0$, $\hat{M}_0=\frac{1}{4}$ need be examined, 
since all others are at least massless. 

For $\hat{M}_0=0$ the corresponding states have vanishing momentum 
and non-trivial winding along the temperature cycle. 
The mass of these states vanishes at $R_0\rightarrow 0$ and diverges at $R_0\rightarrow\infty$. 
The \emph{MSDS} point corresponds to the radius $R_0=\frac{1}{4}$ and is tachyon-free. 
As a result, increasing the radius (lowering the temperature) never produces tachyonic instabilities, 
since the masses of these states can only further increase, 
provided of course that the transverse moduli are kept fixed at their \emph{MSDS} values.
For the states $\hat{M}_0=\frac{1}{4}$, the mass contribution (\ref{firstSquare}) 
has a minimum at $R_0=\frac{1}{4}$ (assuming $n^0=\pm 1$), 
which corresponds to the tachyon-free \emph{MSDS} point. Again, keeping all transverse moduli fixed and varying 
only the temperature modulus $R_0$, we never encounter tachyons since 
the most dangerous point for these states is precisely the \emph{MSDS} point, 
which we know to be free of tachyons.

This establishes the existence of a deformation trajectory $R_0\rightarrow \infty$ 
that takes the initial \emph{MSDS} thermal vacuum to a supersymmetric vacuum at zero temperature, 
without encountering tachyonic instabilities. Once the temperature has fallen considerably, 
marginal deformations along the ``transverse moduli'' are free of tachyons. 
As a result, it is possible to further decompactify one or more spatial dimensions 
to obtain higher-dimensional vacua. If one of the decompactified spatial radii 
is the one coupling to the right $R$-symmetry charges, 
one eventually obtains a $d=4$, $\mathcal{N}=8$ supersymmetric vacuum 
at zero temperature.

Finally, it should be noted that the conditions (\ref{TachyonFreeCond}) 
apply only to the case of string vacua of the ``maximally symmetric type''. 
For vacua in which some of the (dynamical) moduli are twisted by orbifolds, 
the previous argument is no longer necessarily restrictive and it is, in fact, 
possible to construct orbifold \emph{MSDS} vacua that are tachyon-free 
with respect to any deformation in the ``transverse'' moduli space.
We construct such tachyon-free vacua in Section \ref{twisted}.

%%%%%%%%%%%%%%%%%%%%%%%%%%%%%%%%%%%%%%%%%%%%%%%%%

\subsection{Connection to tachyon-free Type II thermal models}

It is interesting to observe that the maximally symmetric \emph{MSDS} vacua 
are continuously connected to the class of tachyon-free Type II thermal vacua 
constructed in \cite{akpt}. In what follows we present the interpolation between 
the two classes of models. 

We take the representation (\ref{NewLattice}) for the $\Gamma_{(8,8)}$ lattice, and consider 
large marginal deformations along which 
the $T^6$ torus decompactifies. 
In this limit, only the zero orbits survive:
\be
	(m_2,n_2)=(m_3,n_3)=(\tilde{m}_4,\tilde{n}_4)=(m'_2,n'_2)=(m'_3,n'_3)=(M',N')=(0,0),
\ee
which effectively decouples the full lattice into the $(2,2)$ and $(6,6)$ sublattices. 
with their contribution reducing to a volume factor $V_6$.
The original lattice (\ref{NewLattice}) reduces into the product of two $\Gamma_{(1,1)}$ lattices, 
each ``thermally'' coupled to the left-moving and right-moving $R$-symmetry charges, respectively:
$$
V^0=\tilde m^0 + \tau n^0,~~~~~~V^1=\tilde m^1 +\tau n^1,
$$
\be
	-\frac{1}{2\tau_2}\left(|V^0|^2+|V^1|^2\right)
+i\pi\left[\tilde m^1 n^1+\bar{a}\tilde m^1+\bar{b}n^1+\tilde{m}^0{n}^0+a\tilde{m}^0+b{n}^0  \right].
\ee
One may then shift $V^1\rightarrow V^1 + V^0$ in order to obtain 
the thermal lattice of \cite{akpt} at the fermionic point in moduli space:
\be\label{AngelantonjModel}
	\Gamma_{(8,8)}\rightarrow \frac{ V_{6}}{(2\tau_2)^4}
\sum\limits_{\tilde m^0,n^0,\tilde{m}^1,{n}^1\in\Z}{e^{-\frac{\pi}{2\tau_2}\left(|V^0|^2+|V^0+V^1|^2\right)+
	i\pi\left[\tilde{m}^0(a+\bar{a})+ n^0(b+\bar{b})+(\tilde m^1 n^1+\bar{a}\tilde m^1+\bar{b}n^1)+(\tilde m^1n^0-\tilde{m}^0 n^1)\right]}}.
\ee
In this form the $X^{0}$ cycle couples to the total spacetime fermion number $F_L+F_R$, 
while the $X^{1}$ cycle couples only to the right-moving fermion number $F_R$. 

Following the discussion of the previous section on the thermal interpretation 
of the maximally symmetric \emph{MSDS} models, we can identify the Euclidean time direction 
along the $X^{0}$ circle, and then its radius  
determines the inverse temperature $\beta=2\pi R_0$. 
The chemical potentials are given by $\mu=2 \tilde \mu=i$. 
Freezing the chemical potentials at these special values, the models were shown to be tachyon-free
under radial deformations and to be characterized by thermal duality symmetry in \cite{akpt}.
Notice that they satisfy the conditions (\ref{TachyonFreeCond}), derived in the previous section, for such a thermal model to be
tachyon-free under arbitrary marginal deformations.
A four-dimensional thermal model of this type can be obtained if we compactify 
the two longitudinal dimensions, and re-compactify 3 directions associated with the
$(6,6)$ sub-lattice. The cosmology of these models is under investigation \cite{BKPPT}.

%%%%%%%%%%%%%%%%%%%%%%%%%%%%%%%%%%%%%%%%%%%%%%%%%%
%%%%%%%%%%%%%%%%%%%%%%%%%%%%%%%%%%%%%%%%%%%%%%%%%%%%%%%%%%%%%%

\section{The tachyon-free Hybrid models}\label{Hybrid}

In the previous section,
we developed the general framework needed to describe 
the marginal deformations of the \emph{MSDS} models. 
Our analysis focused on the cases of the maximally symmetric Type II and Heterotic models. 
Even though these points are non-singular, arbitrary small marginal deformations generically
produce tree-level tachyonic instabilities. In the remaining 
two sections of this paper, we present classes of non-singular models that remain
tachyon-free under arbitrary marginal deformations.

It was found in Section \ref{stability} (see also Appendix \ref{conditionsCheck}) that in order to construct thermal-like models, 
which are tachyon-free under all\footnote{With the exception 
of the frozen chemical potentials that appear in the thermal ensemble of the models.} 
possible deformations of the dynamical moduli, 
the compactification lattice should admit a factorization 
involving a $(1,1)$-lattice factor 
coupled to the left-moving fermion number $F_L$ only, in a suitable basis:
\be\label{Factorization}
	\Gamma_{(d,d)}[^{a,~\bar{a}}_{b,~\bar{b}}]=\Gamma_{(1,1)}[^{a}_{b}](R_0)\otimes
\Gamma_{(d-1,d-1)}[^{\bar{a}}_{\bar{b}}](G_{IJ},B_{IJ})~.
\ee
This $(1,1)$ factor is associated with the Euclidean time circle.
In this section we construct a large class of such thermal models and analyse their stability 
under marginal deformations. They are based on asymmetric (freely acting) orbifold compactifications of two-dimensional
\emph{Hybrid MSDS} models to one dimension. In the Hybrid models, 
spacetime supersymmetry arises from the left-moving side of the string. 
The right-moving supersymmetries are broken at the string scale 
and are replaced by the \emph{MSDS} structure.
The full supersymmetry will be spontaneously broken by the asymmetric (freely acting) orbifold compactification 
to one dimension. 

\subsection{The Hybrid models}
We can construct a two-dimensional Hybrid model as follows. 
The $24$ left-moving fermions are split into two groups of $8$ and $16$,
described in terms of $SO(8)$ and $E_8$ characters respectively.
The $16$ fermions arise via fermionization of the $8$ left-moving internal
coordinates: $i\partial X^I_L=y^Iw^I(z)$, $I=1\dots 8$. 
As in the Type II \emph{MSDS} models, the right-moving fermions are described in terms of 
$SO(24)$ characters. 
The partition function is given by:
$$
 Z = {V_2\over (2\pi)^2}~\int_\F {d^2\tau\over
4({\rm Im}\tau)^{2}} 
~\left[\frac 12 \sum_{a,b} (-)^{a+b}~ {\Th ab^{4} \over \eta^{4}}\right]~
\left[\frac 12 \sum_{\gamma,\delta} ~ {\theta[^{\gamma}_{\delta}]^{8} \over \eta^{8}}\right]~
\left[\frac 12 \sum_{\bar a, \bar b} (-)^{\bar a+\bar b} ~ {\bar \theta[^{\bar a}_{\bar b}]^{12} \over \bar \eta^{12}}\right]
$$
\be
={V_2\over (2\pi)^2}~\int_\F {d^2\tau\over
4({\rm Im}\tau)^{2}}~{1 \over \eta^8}~\Gamma_{E_8}(\tau)~\left(V_8 - S_8\right)~\left(\bar V_{24}-\bar S_{24}\right).
\ee
In the last line we have expressed the partition function in terms of the $SO(8)$ and
$SO(24)$ characters and of the chiral $E_8$ lattice.
This is a $(4,0)$ supersymmetric model with respect to ordinary supersymmetry, with an
\emph{MSDS}-symmetric anti-holomorphic sector
with respect to the \emph{MSDS} symmetry of \cite{massivesusy,reducedMSDS}.
Both the left- and right-moving characters are by themselves modular invariant. 
Because of supersymmetry, the partition function is identically
zero. 
There are $24 \times 8$ massless bosons and $24 \times 8$ massless fermions,  
arising in the $V_8~ \bar V_{24}$ and $S_8~ \bar V_{24}$ sectors respectively. 
A large class of chiral orbifolds that preserve the right-moving \emph{MSDS} structure can be constructed following
\cite{reducedMSDS}.

It is interesting to note that
the Hybrid model can be obtained 
from the maximally symmetric Type II \emph{MSDS} model by performing 
an asymmetric $\Z_2$ orbifold $X^I_L\rightarrow X^I_L+\pi$, 
which shifts the transverse left-moving coordinates. 
The orbifold projection breaks the left-moving gauge group $H_L=[SU(2)_L]_{k=2}^8$ of the 
maximally symmetric \emph{MSDS} model down to an abelian $U(1)_L^8$, 
whereas $H_R=[SU(2)_R]_{k=2}^8$ remains unbroken. 
The latter is spontaneously broken to $U(1)_R^8$ as soon as one 
deforms the model away from the fermionic point. 

To exhibit the model as a Type II asymmetric orbifold compactification to
two dimensions, we write the partition function as follows 
$$
Z= {V_2\over (2\pi)^2}~\int_\F {d^2\tau\over
4({\rm Im}\tau)^{2}}~ Z_{\rm Hybr.},
$$
\be\label{HybridPart}
Z_{\rm Hybr.}=\frac{1}{\eta^{12}\bar\eta^{12}}~\frac{1}{2}\sum\limits_{a,b=0,1}
{(-)^{a+b}~\theta[^a_b]^4}~\frac{1}{2}\sum\limits_{\bar{a},\bar{b}}
{(-)^{\bar{a}+\bar{b}}~\bar\theta[^{\bar{a}}_{\bar{b}}]^{4}~\Gamma_{(8,8)}[^{\bar{a}}_{\bar{b}}]}~.
\ee
We identify the internal $(8,8)$ lattice at the fermionic point with:
\be\label{latticeFermPoint}
\Gamma_{(8,8)}[^{\bar{a}}_{\bar{b}}]=\Gamma_{E_8}\times\bar\theta[^{\bar{a}}_{\bar{b}}]^8
=\frac{1}{2}\sum\limits_{\gamma,\delta=0,1}{\theta[^\gamma_\delta]^{8}}~\bar\theta[^{\bar{a}}_{\bar{b}}]^8.
\ee
This is nothing but the asymmetrically half-shifted $(8,8)$ lattice analysed in detail 
in Section \ref{88Lattice}. The summation over the left-moving charges $(\gamma,\delta)$ imposes that
the winding numbers $(m,n)$ coupling to them be even integers. Since the Hybrid $(8,8)$
lattice couples to the right-moving fermion number $F_R$ only, it suffices to infinitely
deform a single radial modulus in order to recover a maximally supersymmetric model. 

\subsection{Asymmetric orbifold compactification to one dimension}
Next we compactify one of the longitudinal directions on a circle, whose radius
we denote by $R$. 
In particular, we consider an asymmetric orbifold obtained  
by modding out with $(-1)^{F_L}\,\delta$, where $\delta$ is an order-$2$ shift along the compact circle and
$F_L$ is the left-moving fermion number. 
We will eventually be interested in the thermal interpretation of the model, 
identifying the circle with the Euclidean time direction.
Throughout we use the following notations 
associated with the $\Gamma_{(1,1)}$ lattice:
$$
\Gamma_{(1,1)} (R) = ~\frac{R}{\sqrt{\tau_2}}\sum_{\tilde m , n} 
e^{-\pi \frac{R^2}{\tau_2} |\tilde m + n \tau |^2}
=~ \sum_{m,n} \Gamma_{m,n} \,,             
$$ 
\be
\label{lattice}
\Gamma_{m,n} = q^{\frac{1}{2} p_{ L}^2 }\, \bar q ^{\frac{1}{2} p_{ R}^2}\;,\quad   
q=e^{2\pi i \tau}\;,\quad \mbox{and}\quad      p_{ L,R} = {1 \over \sqrt{2}}\left(\frac{m}{R} \pm n R\right)\,.
\end{equation}

The partition function is given by \cite{akpt}
$$
{Z \over V_1} = \int_{\cal F} \frac{d^2 \tau}{8\pi({\rm Im}\tau)^{3/2}}~(\bar V_{24}- \bar S_{24}) \, 
~{\Gamma_{E_8}\over \eta^8} 
$$
\begin{equation}
\times \sum_{m, n}
\left(V_8 \, \Gamma_{m,2n} + O_8 \, \Gamma_{m +\frac{1}{2},2n +1} - S_8 \, 
\Gamma_{m +\frac{1}{2} , 2 n} -C_8\, \Gamma_{m, 2n +1} \right).
\label{parth}
\end{equation}
The deformation affects the left movers and breaks the $(4,0)$ supersymmetry spontaneously.
For generic values of $R$, the only massless states arise in the $V_8\bar V_{24}$ sector. 
The initially massless fermions in the $S_8\bar V_{24}$ sector are now massive.
Notice the appearance of the $O_8$ sector, which contains the left-moving NS vacuum. 
Here, as in \cite{akpt}, 
it carries non-trivial
momentum and winding numbers along the circle. The right-moving sector, on the other hand, 
begins at the massless level and so
the model remains tachyon-free for all values of $R$. 
The model is $T$-duality invariant, where now $R \to 1/2R$
and as usual the $S_8$ and $C_8$ sectors get interchanged. 

The lowest mass in $O_8\bar V_{24}$ sector is given by
\be
2m_{O\bar V}^2=\left({1\over \sqrt{2}R}-\sqrt{2}R\right)^2. 
\ee  
So from this sector we get additional massless states at the fermionic point, $R=1/\sqrt{2}$. These
states induce non-analyticity in $Z$, precisely at this point. As we shall see, the partition function
is finite at this point, but its first derivative is discontinuous. The second derivative diverges. Thus we get a
first order phase transition. In the 
higher-dimensional model of \cite{akpt}, the transition is much milder. 
At $R=1/\sqrt{2}$ there is a further enhancement of the gauge symmetry:
\be
U(1)_L \times U(1)_R \to SU(2)_L \times U(1)_R~.
\ee
This can be understood as follows. Due to the asymmetric nature of the orbifold,
which involves a twist by $(-1)^{F_L}$, the $U(1)$ current algebra of $i\partial X$ 
becomes extended to $SU(2)_{k=2}$. The boundary conditions of the longitudinal spacetime fermion 
$\psi$ become correlated to those of the $\chi,\omega$ 
fermions arising from the fermionization of $\partial X$ at the fermionic radius $R=1/\sqrt{2}$. 
Similar phase transitions have been found to occur in the context
of non-critical Heterotic strings in two dimensions in \cite{DLS}.
Two-dimensional non-critical strings, as well as their matrix model
descriptions, also exhibit thermal duality symmetry (see e.g. \cite{Klebanov}).

We now proceed to compute the partition function. 
For $R > 1/\sqrt{2}$, we Poisson re-sum over the momentum
$m$ to obtain
$$
{Z \over V_1} = \int_{\cal F} \frac{d^2 \tau}{8\pi({\rm Im}\tau)^{3/2}}
~(\bar V_{24}- \bar S_{24}) \,{\Gamma_{E_8}\over \eta^8} 
$$
\be
\times \frac 12 \sum_{a,b}~(-)^{a+b}~{R \over \sqrt{\tau_2}}\sum_{\tilde m, n}
e^{-\frac{\pi R^2}{\tau_2}|\tilde m +n\tau|^2}
~ (-)^{\tilde m a+nb+\tilde m n}~ {\Th ab^{4} \over \eta^{4}}\,.
\label{parth1}
\end{equation}
The last line in the integrand of equation (\ref{parth1}) 
exhibits the asymmetric nature of the (freely acting) orbifold in
terms of a $\Gamma_{(1,1)}$ lattice, 
``thermally'' coupled to the left movers.
Explicitly, the last line is given by 
\be\label{Thermalcharacters}
\Gamma_{(1,1)}(R)~(V_8-S_8)-{1\over \eta^{4}}
\left(\Gamma[^1_1]~\theta_3^{4}-\Gamma[^1_0]~\theta_4^{4}-\Gamma[^0_1]~\theta_2^{4}\right), 
\ee
where we introduced the notation for the shifted latices
\be
\Gamma[^h_{\tilde g}]={R \over \sqrt{\tau_2}}\sum_{\tilde m, n}
e^{-\frac{\pi R^2}{\tau_2}|(2\tilde m+\tilde g) +(2n+h)\tau|^2}.
\ee
The untwisted piece in (\ref{Thermalcharacters}) vanishes
due to the initial left-moving supersymmetries.
In total we get:
\be
{Z \over V_1}=-\int_{\cal F} \frac{d^2 \tau}{8\pi({\rm Im}\tau)^{3/2}}~
{1\over \eta^{4}}\left(\Gamma[^1_1]~\theta_3^{4}-\Gamma[^1_0]~\theta_4^{4}-\Gamma[^0_1]~\theta_2^{4}\right)~
{\Gamma_{E_8}\over \eta^8}~(\bar V_{24}- \bar S_{24}).
\label{parth2}
\ee

We will compute the integral by mapping it to the strip. 
Notice that for very small $R$, the various terms 
appearing in the lattice sum over $(\tilde m, n)$ 
in (\ref{parth2}) (or (\ref{parth1})) do
not give convergent integrals by themselves. 
To be able
to exchange the order of summation and integration in the small $R$ case, 
we should Poisson re-sum over the
winding number $n$ in (\ref{parth}). Then we obtain a similar 
expression as in (\ref{parth1}), with $R\to 1/2R$. 
This is essentially $T$-duality. In Appendix \ref{Duality}, we explicitly demonstrate the action of
$T$-duality by performing a double Poisson re-summation (see \ref{DoublePoissonForm} in Appendix \ref{DoublePoisson}) on the thermally shifted 
$\Gamma_{(1,1)}$ lattice 
appearing in (\ref{parth1}).
Alternatively, we can compute the integral for $R > 1/\sqrt{2}$,
and then apply $T$-duality to the result in order to find
the answer for $R < 1/\sqrt{2}$ \cite{DLS}.   

Before we proceed, we check how modular transformations
act on the various terms in the integrand of (\ref{parth2}).
The right-moving character $\bar V_{24}-\bar S_{24}$
is modular invariant.
Under $\tau \to \tau+1$ the rest transform as follows
$$
\Gamma[^1_1]~~\leftrightarrow ~~\Gamma[^1_0],~~~~~
\,\,\,\, \Gamma[^0_1]~~\leftrightarrow~~\Gamma[^0_1]
$$
\be
{\theta_3^4~\Gamma_{E_8} \over \eta^{12}}~~ \leftrightarrow ~~ 
-{\theta_4^4~\Gamma_{E_8} \over \eta^{12}}~,
\,\,\,~~~~~~ {\theta_2^4~ \Gamma_{E_8} \over \eta^{12}}~~
\leftrightarrow~~ {\theta_2^4~\Gamma_{E_8} \over \eta^{12}}~.
\ee
Under $\tau \to -1/\tau$,
$$
\Gamma[^1_1]~~\leftrightarrow~~ \Gamma[^1_1],~~~~~
\,\,\,\, \Gamma[^1_0]~~\leftrightarrow~~ \Gamma[^0_1]
$$
\be
{\theta_3^4~\Gamma_{E_8} \over \eta^{12}}~~ \leftrightarrow ~~ 
{\theta_3^4 ~\Gamma_{E_8} \over \eta^{12}},\,\,\,~~~~~~ 
{\theta_4^4~\Gamma_{E_8} \over \eta^{12}}~~\leftrightarrow~~ {\theta_2^4~\Gamma_{E_8} \over \eta^{12}}~.
\ee
So modular transformations act as permutations, 
where the shifted lattices and the accompanying left-moving theta functions 
are permuted in exactly the same way. 

Now consider the three sets of integers appearing in the various
orbits of the shifted lattices:
\be
(2\tilde m+1,2n+1), \,\,\, (2\tilde m, 2n+1), \,\,\, (2\tilde m+1, 2n).
\ee
In each set we can write the pair of integers as $(2\tilde k +1)(p,q)$, 
where the greatest common divisor is odd and
$p$ and $q$ are relatively prime. 
Among all sets, we generate in this way all pairs of relatively prime integers. 
For each such pair, we can find a unique modular transformation that maps 
$(p,q)\to (1,0)$, and the fundamental region $\F$ to
a region in the strip \cite{FtoStrip}. 
Any two such regions are non-intersecting and their union makes up the entire strip. 
Since the relevant modular transformation maps $(p,q)\to (1,0)$, 
the corresponding orbit gets mapped to an orbit in 
the $\Gamma[^0_1]$ shifted lattice. 
So the accompanying theta function gets permuted to $\theta_2$. 
The partition function for $R>1/\sqrt{2}$, can be written as
follows
\be\label{strip}
{Z \over V_1}=2R\sum_{\tilde k=0}^{\infty}\int_{||} \frac{d^2 \tau}{8\pi({\rm Im}\tau)^{2}}~
e^{-(2\tilde k+1)^2{\pi R^2\over \tau_2}}~{\theta_2^{4}\over \eta^{12}}~\Gamma_{E_8}(\tau)
~(\bar V_{24}-\bar S_{24}).
\ee
 
The right-moving \emph{MSDS} structure allows us 
to compute the integral exactly.
Since
$\bar V_{24}-\bar S_{24}=24$, level matching implies 
that the only non-vanishing
contributions are those of the massless level 
\be
{Z \over V_1}=2R\sum_{\tilde k=0}^{\infty}\int_{||} \frac{d^2 \tau}{8\pi({\rm Im}\tau)^{2}}~
e^{-(2\tilde k+1)^2{\pi R^2\over \tau_2}}~ (16 \times 24).
\ee 
In the parenthesis we get the multiplicity of 
the massless level of the initially supersymmetric model.
In all we have
\be
{Z \over V_1}=(16 \times 24)~{1 \over 2\pi^2 R}~\sum_{\tilde k=0}^{\infty}{1\over (2\tilde k +1)^2}
=(16 \times 24)~{1 \over 16 R}={24 \over R}.
\ee

The result above is valid for $R > 1/\sqrt{2}$. 
To get the answer for $R < 1/\sqrt{2}$, we impose the $R \to 1/2R$ duality.
We get 
\be
{Z \over V_1}=24 \times (2R).
\ee
The formula valid for all $R$ is
\be \label{partfinal}
{Z \over V_1}=24\times \left(R + {1 \over 2R}\right)-24 \times \left| R - {1 \over 2R}\right|.
\ee
The result is manifestly invariant under the $R\to 1/2R$ duality, and contains the expected non-analyticity
induced by the extra massless states at the dual fermionic point $R=1/\sqrt{2}$.
In fact, the non-analytic part of 
$Z$ can be written as
\be
-24~|m_{O\bar V}(R)|
\ee
and can be understood as follows. Consider the contribution from the 
massless states in the $O_8\bar V_{24}$ sector to $Z$, 
equation (\ref{parth}), near the extended symmetry point. 
There are 24 complex (or 48 real) such scalars. Their contribution is given by
\be
48 \times \int_1^{\infty}{dt \over 4\pi t}~t^{-1/2}~e^{-\pi m^2 t}
=48~ |m|~{1\over 4\sqrt{\pi}}\int_{\pi m^2}^{\infty}{dy \over y}~y^{-1/2} e^{-y}. 
\ee 
In the limit $m\to 0$, the leading contribution is
\be
48~|m|~{1\over 4\sqrt{\pi}}\Gamma\left(-{1 \over 2}\right)=-24~|m|.
\ee
From equation (\ref{partfinal}), we see that the first derivative of
the partition function is discontinuous and so we have a first order
phase transition which connects the two dual phases.
When enough additional dimensions decompactify, the transition 
becomes a higher order one and it will be essentially smooth.

\subsubsection{Thermal interpretation}\label{thermalHybrid} 
We now identify the additional compact cycle with the Euclidean time direction.
To analyse the thermal interpretation of the model, we begin with equation 
(\ref{strip}). We may rewrite it using (\ref{latticeFermPoint}) as follows
\be
{Z \over V_1}=2R_0\sum_{{\tilde m}^0=0}^{\infty}~\int_{||} \frac{d^2 \tau}{8\pi({\rm Im}\tau)^{2}}~
e^{-(2{\tilde m}^0+1)^2{\pi R_0^2\over \tau_2}}~{\theta_2^{4}\over \eta^{12}}~
\frac{1}{2~\bar \eta^{12}}\sum\limits_{\bar{a},\bar{b}}
{(-)^{\bar{a}+\bar{b}}~\bar\theta[^{\bar{a}}_{\bar{b}}]^{4}~\Gamma_{(8,8)}[^{\bar{a}}_{\bar{b}}]}~.
\ee
This expression exhibits the model as
an asymmetric orbifold compactification of Type II theory to one dimension.
Since the Euclidean time cycle is factorized,
the above formula remains valid for arbitrary deformations of the
``transverse'' moduli associated with the $(8,8)$ internal lattice.
The integral is finite but harder to compute analytically, once we deform away from the \emph{MSDS} point.

The one-loop partition function is also written as
$$
{Z \over V_1}=R_0\sum_{{\tilde m}^0 \ne 0}^{\infty}~\int_{||} \frac{d^2 \tau}{8\pi({\rm Im}\tau)^{2}}~
e^{-{\pi({\tilde m}^0R_0)^2\over \tau_2}}~{1\over 2~\eta^{12}}~
\sum\limits_{a,b}{(-)^{a+b}~\theta[^a_b]^{4}~(-)^{{\tilde m}^0 a}}
$$
\be
\times~\frac{1}{2~\bar \eta^{12}}\sum\limits_{\bar{a},\bar{b}}
{(-)^{\bar{a}+\bar{b}}~\bar\theta[^{\bar{a}}_{\bar{b}}]^{4}~\Gamma_{(8,8)}[^{\bar{a}}_{\bar{b}}]}~.
\ee
As the integral is over the strip, we can
easily infer the complete 
spacetime partition function. It is given by
the right-moving fermion index
\be
\mathcal{Z}=\textrm{tr}{\left[~e^{-\beta H}~
(-1)^{F_R}\right]}~,
\ee
which now, because of the phase transition, 
is strictly valid for $R_0>1/\sqrt{2}$.
The trace is over the Hilbert space of the 
initially supersymmetric $(4,0)$ model and $\beta=2\pi R_0$.
For $R_0 < 1/\sqrt{2}$, we get a similar expression but with
$\beta=\pi/R_0$. Thus the system at small 
radii is again effectively cold.
Since there is no unique expression valid
for all values of the radius $R_0$, the underlying
duality symmetry of the model $R_0 \to 1/2R_0$ is not manifest in the ``thermal'' trace. The fundamental
object is, rather, the Euclidean path integral, which is valid
for all temperatures and manifestly exhibits the stringy thermal duality
of the model.

The $(8,8)$ internal lattice can be written in the Hamiltonian form
in terms of the left- and right-moving momenta, which are explicitly given by
\be
P_{L,R}^I=\frac{1}{\sqrt{2}}G^{IJ}\left(\hat{m}_J+(B_{JK}\pm G_{JK})n^K\right),
\ee
where $\hat{m}_J\equiv m_J+\frac{1}{2}(\bar{a}+n^1)\delta_{1J}$.
The momentum along the $X^1$ direction is half-shifted due to 
the coupling to the right-moving fermion number $F_R$.
Using these expressions, we can exhibit the trace as a conventional thermal
ensemble, which is further deformed by chemical potentials associated with ``gravito-magnetic''
fluxes, as in equation (\ref{ThermalInterpretationMAX}): 
\be
\mathcal{Z}(\beta,\mu, \tilde \mu) = \textrm{tr}\left[~e^{-\beta H}~
e^{2\pi (\mu\,\hat m_1-\tilde \mu\, n^1)} \right]~.
\ee
Here $\mu=i$, $\tilde \mu=\frac{i}{2}$ are the imaginary chemical potentials 
coupled to the $U(1)$ charges
\be
	\hat m_1=m_1+\frac{1}{2}(\bar{a}+n^1), ~~~~n^1.
\ee
As we anticipated in Section \ref{stability}, freezing these particular 
values of the chemical potentials, or equivalently, 
when the $(1,1)$ and $(8,8)$ lattices are initially factorized, 
the model is free of tachyonic instabilities for arbitrary deformations of the
``transverse'' moduli. We verify this explicitly in the following section.
Essentially, the stability is guaranteed by the thermal duality,
$R_0 \to 1/2R_0$, of the
model.

\subsection{Deformations and stability of the thermal Hybrid models}
We next analyse the stability of the thermal Hybrid models
under marginal deformations. 
We show that they are free of tachyonic instabilities for any 
deformation in the ``transverse'' moduli space.  
The integrand of the partition function can be fully decomposed 
in terms of the $SO(8)$ characters as follows
$$
	\sim \frac{1}{\eta^8\bar\eta^8}\sum\limits_{m_0,n_0}{\left(V_8 \Gamma_{m_0,2n_0}^{(1,1)}
+O_8 \Gamma_{m_0+\frac{1}{2},2n_0+1}^{(1,1)}-S_8 
\Gamma_{m_0+\frac{1}{2},2n_0}^{(1,1)}-C_8 \Gamma_{m_0,2n_0+1}^{(1,1)}\right)}
$$
\be
	\times\sum\limits_{M,N}{\left(\bar{V}_8\Gamma_{M,2N}^{(8,8)}
+\bar{O}_8 \Gamma_{M+\frac{1}{2},2N+1}^{(8,8)}-\bar{S}_8\Gamma_{M+\frac{1}{2},2N}^{(8,8)}
-\bar{C}_8\Gamma_{M,2N+1}^{(8,8)}\right)},
\ee
where $(m_0,n_0)$ are the momenta and windings along the Euclidean time circle. 
The decomposition in the second line for the $\Gamma_{(8,8)}$ lattice is given 
in terms of the momentum and winding modes $(M,N)$ 
along the cycle that couples to the right-moving $R$-symmetry charges. 
The latter can be obtained 
by Poisson re-summing the Lagrangian lattice (\ref{NewLattice}). 

We focus on the $O_8 \bar{O}_8$ sector since this is the only sector dangerous of producing tachyons. 
The lowest mass states in this sector are entirely determined 
by the $\Gamma_{(1,1)}\oplus\Gamma_{(8,8)}$ lattice contributions. 
The $\Gamma_{(1,1)}(R_0)$ contribution to the mass formula 
is given in terms of the left- and right-moving momenta along the Euclidean time circle:
\be
	\frac{1}{2}P_{L,R}^2 = \frac{1}{4}\left( \frac{m_0+\frac{1}{2}(a+n_0)}{R_0}\pm n_0 R\right)^2.
\ee
In the $O\bar{O}$ sector, the windings $n_0$ are odd. 
Notice that the $(1,1)$ lattice is uncoupled from the $\Gamma_{(8,8)}$ lattice 
and so
\be
	\min\left|\frac{1}{2}(P_L^2-P_R^2)\right|=\min\left|\left(m_0+\frac{1}{2}n_0\right)n_0\right|=\frac{1}{2}.
\ee 
This is clearly just enough to produce a state that is massless \emph{at least} holomorphically 
(or antiholomorphically). This observation is independent of the value of the radius $R_0$ 
or of the other moduli $G_{IJ}, B_{IJ}$ of $\Gamma_{(8,8)}$. 
It crucially depends, however, on the factorization $\Gamma_{(1,1)}\oplus\Gamma_{(8,8)}$ 
into lattices with independent couplings to the left and right $R$-symmetry charges, respectively. 

There is another way to see the stability as a property 
of the factorization of the coupling on the left and right $R$-symmetries. 
To illustrate this, consider the two lattices $\Gamma_{(1,1)}[^a_b]$, $\Gamma_{(8,8)}[^{\bar{a}}_{\bar{b}}]$. 
The difference $\Delta_{L}-\Delta_R\equiv\frac{1}{2}(P_L^2-P_R^2)$ between the left- and right-moving 
conformal weights of a lattice is always independent of the deformation parameters, 
as shown in (\ref{levelmatching}). 
Therefore, one may calculate it at the fermionic point where the  
representation (\ref{latticeFermPoint}) in terms of $\vartheta$-functions is available. 
In this case one may further decompose the lattice part of 
the ``dangerous'' sector (\ref{dangerousSector}) 
in terms of $SO(2)\times SO(16)$ characters:
$$
	\left(O_2\bar{V}_2+V_2\bar{O}_2\right)\times\left(O_{16}\bar{V}_{16}+C_{16}\bar{V}_{16}\right).
$$
The two lattices are factorized and one may calculate the (partial) 
difference in the conformal weights of the \emph{lowest lying states} independently:
$$
	O_2\bar{V}_2~:~(\Delta_L-\Delta_R)_{(1,1)}=-\frac{1}{2}~~,
~~	V_2\bar{O}_2~:~(\Delta_L-\Delta_R)_{(1,1)}=+\frac{1}{2},
$$
$$
	O_{16}\bar{V}_{16}~:~(\Delta_L-\Delta_R)_{(8,8)}=-\frac{1}{2}~~,
~~	C_{16}\bar{V}_{16}~:~(\Delta_L-\Delta_R)_{(8,8)}=+\frac{1}{2}.
$$
This implies that the difference in the conformal weights for \emph{each of the two lattices separately} is
\be\label{levelunmatch}
	|\Delta_L-\Delta_R|\equiv \left|\frac{1}{2}(P_L^2-P_R^2)\right| = \frac{1}{2},
\ee
independently of the deformations. For Type II theories, 
where the lowest possible weight is $(-\frac{1}{2})$ for both the left- and the right-moving NS-vacuum, 
this forces the left- or right-moving sector to be at least massless and, 
thus, guarantees the absence of \emph{physical} tachyons from the spectrum 
for any deformation that preserves the factorization (\ref{Factorization}).

Explicitly, the mass of the lowest $O\bar{O}$ states is given by
\be
	m^2_{O\bar{O}} = \frac{1}{2}\left(\frac{1}{2R_0}-R_0\right)^2+m^2_{(8,8)}-\frac{1}{2},
\ee
where $m^2_{(8,8)}$ is the $(8,8)$-lattice contribution:
\be
	m^2_{(8,8)} \equiv \frac{1}{2}(P_{L,(8,8)}^2+P_{R,(8,8)}^2) \geq |\Delta_{L}-\Delta_{R}|_{(8,8)}.
\ee
In view of (\ref{levelunmatch}), $m_{(8,8)}^2\geq \frac{1}{2}$ for any deformation and, 
thus, the physical states in the $O\bar{O}$-sector are at least massless.

We conclude this section by describing some simple deformation limits. 
Assuming the form (\ref{latticeFermPoint}) 
for the $\Gamma_{(8,8)}$ lattice at the fermionic point, we obtain the following.
\begin{itemize}
	\item In the limit of infinite radius $R_0\rightarrow\infty$, 
corresponding to zero temperature, the left-moving $O$ and $C$ sectors 
become infinitely massive and decouple from the spectrum, 
whereas the fermionic states $S_{8}\bar{V}_{24}$ become massless and one recovers 
the supersymmetric $(4,0)$ Type IIB Hybrid model.
	\item In the zero radius limit $R_0\rightarrow 0$, 
the left-moving $O$ and $S$ sectors decouple, 
while the fermionic states that become massless are now $C_{8}\bar{V}_{24}$. 
One, therefore, recovers the supersymmetric $(4,0)$ Type IIA Hybrid model.
	\item At the self dual radius $R_0\rightarrow 1/\sqrt{2}$ 
one obtains extra massless states, which is the signal of enhanced gauge symmetry. 
As previously discussed, this is the enhanced $[SU(2)_L\times U(1)_R]$ gauge symmetry 
which originates from the worldsheet $[SU(2)_L]_{k=2}\times U(1)_R$ current algebra.
	\item Another interesting case is the limit in which the $(8,8)$ lattice asymptotically 
decouples from the right-moving $R$-symmetry charges, at zero temperature. 
This limit asymptotically restores the 4 right-moving supersymmetries. 
The relevant modulus corresponds to the supersymmetry breaking scale $M$.
	\item Deforming the $(8,8)$ lattice along the radial directions that are orthogonal to the 
$M$-modulus, one obtains new non-compact dimensions in the decompactification limit. 
In these limits, new stable vacua may be obtained in $d\geq 1$ spacetime dimensions, 
where supersymmetry is either (asymptotically) present or spontaneously broken 
in the presence of thermal and ``gravito-magnetic'' fluxes.
\end{itemize}

%%%%%%%%%%%%%%%%%%%%%%%%%%%%%%%%%%%%%%%%%%%%%%%%%%%%%%%%%%%%%%
%%%%%%%%%%%%%%%%%%%%%%%%%%%%%%%%%%%%%%%%%%%%%%%%%%%%%%%%%%%%%%

\section{Tachyon-free \emph{MSDS} orbifold models in two dimensions}\label{twisted}

In this section we present a second class of tachyon-free Type II \emph{MSDS} theories, 
constructed as $\Z_2\times\Z_2$ orbifolds of the maximally symmetric 
\emph{MSDS} models. The possible $\Z_2^N$ orbifolds preserving the 
\emph{MSDS} structure were classified in \cite{reducedMSDS}. 
Here we will be interested in orbifolds of the internal $\hat{c}=8$ compact CFT that realizes 
a chiral (antichiral) \emph{MSDS} algebra in both the holomorphic and antiholomorphic sectors. 
Initially, the target space will be taken to be a 2-dimensional Euclidean or Minkowski background. 
The choice of the time direction will determine the interpretation of the model along the lines 
described in the previous sections. 
The desired vacuum should have the property that all 
extra massless states become massive as one deforms away from the
extended symmetry point. Essentially, the orbifolds project out
the dangerous moduli that lead to tachyonic instabilities in the maximally symmetric models. 
Extended symmetry points are associated to an affine Lie algebra, 
whose Weyl reflections give rise to the various $T$-dualities of the theory. 
As a result, one expects the masses of extra physical states in these stable vacua to be invariant under 
the duality group with a minimum at the self-dual point, 
when variations of the VEVs of the various moduli are considered.

In this section we will combine $\Z_2$-orbifold shifts and twists \cite{GKR} 
in order to construct \emph{MSDS} vacua stable under all possible deformations of 
the remaining (fluctuating) moduli. Both cold and thermal tachyon-free vacua
can be constructed. We first discuss the construction of the models and comment on 
their stability under marginal deformations. Then we discuss various deformation limits, 
including their (continuous) connection with higher-dimensional supersymmetric vacua 
and, in particular, with conventional four-dimensional superstring vacua. 

\subsection{Tachyon-free \emph{MSDS} orbifold models: general setup}

We now consider $\Z_2\times\Z_2'$ asymmetric orbifold twists 
of the maximally symmetric Type II \emph{MSDS} model acting on four 
of the internal coordinates as follows:
$$
	X^I_L(z)\rightarrow (-)^{g} X^I_L(z),
$$
for the left-moving internal bosons and
$$
	X^I_R(\bar{z})\rightarrow (-)^{g+g'} X^I_R(\bar{z}),
$$
for the right-moving ones, with $g,g'\in\{0,1\}$ and for $I=5,6,7,8$. 
Furthermore, we introduce independent asymmetric $\Z_2^{(1)}\times\Z_2^{(2)}$ shifts 
on all 8 internal coordinates as follows:
$$
	X^I_L(z)\rightarrow X^I_L(z)+\pi G_1~~,~~\textrm{for}~~I=1,2,3,4
$$
$$
	X^I_R(\bar{z})\rightarrow X^I_R(\bar{z})+\pi G_2~~,~~\textrm{for}~~I=1,2,3,4
$$
$$
	X^J_L(z)\rightarrow X^J_L(z)+\pi G_2~~,~~\textrm{for}~~I=5,6,7,8
$$
$$
	X^J_R(\bar{z})\rightarrow X^J_R(\bar{z})+\pi G_1~~,~~\textrm{for}~~I=5,6,7,8.
$$
The full modular invariant partition function of the model is given by
$$
Z_{Twisted}=\frac{1}{2^4\eta^{12}\bar\eta^{12}}\sum\limits_{h,g,h',g'}
\sum\limits_{a,b}{(-)^{a+b}\theta[^a_b]^2\theta[^{a+h}_{b+g}]\theta[^{a-h}_{b-g}]}
$$
\be
\times \sum\limits_{\bar{a},\bar{b}}{(-)^{\bar{a}+\bar{b}}~\bar\theta[^{\bar{a}}_{\bar{b}}]^2 
\bar\theta[^{\bar{a}+h+h'}_{\bar{b}+g+g'}]\bar\theta[^{\bar{a}-h-h'}_{\bar{b}-g-g'}] }
~\Gamma_{(8,8)}[^{a,\bar{a},h,h'}_{b,\bar{b},g,g'}],
\ee
where the twisted and shifted $\Gamma_{(8,8)}$ lattice is factorized into 
a shifted $\Gamma_{(4,4)}^{(1)}$ lattice and a shifted/twisted $\Gamma_{(4,4)}^{(2)}$ lattice:
\be\label{88TwistedLattice}
	\Gamma_{(8,8)}[^{a,\bar{a},h,h'}_{b,\bar{b},g,g'}] = \frac{1}{2^2}\sum\limits_{H_i,G_i}
\Gamma_{(4,4)}^{(1)}\left[^{a, \bar{a}~;~ H_1, H_2}_{b,\bar{b}~;~ G_1, G_2}\right]~\times~ 
\Gamma_{(4,4)}^{(2)}\left.\left[^{a, \bar{a}~;~ H_1, H_2}_{b,\bar{b}~;~ G_1, G_2}\right|{}^{h, h'}_{g, g'}\right].
\ee
Written at the \emph{MSDS} point, the shifted lattice can be written in terms of free fermion characters
\be\label{shiftedLattice}
	\Gamma_{(4,4)}^{(1)}\left[^{a, \bar{a}~;~ H_1, H_2}_{b,\bar{b}~;~ G_1, G_2}\right] 
= \theta[^{a+H_1}_{b+G_1}]^2 \theta[^{a-H_1}_{b-G_1}]^2 ~\times~ 
\bar\theta[^{\bar{a}+H_2}_{\bar{b}+G_2}]^2 \bar\theta[^{\bar{a}-H_2}_{\bar{b}-G_2}]^2.
\ee
Similarly, the asymmetrically twisted (4,4)-lattice reads
\be\label{twistedLattice}
	\Gamma_{(4,4)}^{(2)}\left.\left[^{a, \bar{a}~;~ H_1, H_2}_{b,\bar{b}~;~ G_1, G_2}\right|{}^{h, h'}_{g, g'}\right] 
= \theta[^{a+H_2+h}_{b+G_2+g}]^2 \theta[^{a-H_2}_{b-G_2}]^2 ~\times~ 
\bar\theta[^{\bar{a}+H_1+h+h'}_{\bar{b}+G_1+g+g'}]^2 \bar\theta[^{\bar{a}-H_1}_{\bar{b}-G_1}]^2~(-)^{h'g'}.
\ee
Its non-vanishing components are collected below:
\be
	\Gamma_{(4,4)}^{(2)}\left.\left[^{a, \bar{a}~;~ H_1, H_2}_{b,\bar{b}~;~ G_1, G_2}\right|{}^{h, h'}_{g, g'}\right] = \left\{
\begin{array}{l}
	\theta[^{a+H_2}_{b+G_2}]^4\bar\theta[^{\bar{a}+H_1}_{\bar{b}+G_1}]^4,~~\textrm{for}~~(h,g)=(h',g')=(0,0) \\
	\theta[^{a+H_2}_{b+G_2}]^4\left(\frac{2\bar{\eta}^3}{\bar\theta[^{1-h'}_{1-g'}]}\right)^2,~~\textrm{for}
\left\{\begin{array}{l}
	(h,g)=(0,0), (h',g')\neq(0,0)~\textrm{and} \\
	~\\
	(H_1,G_1)=(\bar{a},\bar{b})~\textrm{or}~(\bar{a}+h',\bar{b}+g')\\
\end{array}\right. \\
~\\
	\left|\frac{2\eta^3}{\theta[^{1-h}_{1-g}]}\right|^4,~~\textrm{for}
	\left\{\begin{array}{l}
	(h,g)\neq(0,0), (h',g')=(0,0)~\textrm{and} \\
	(H_1,G_1)=(\bar{a},\bar{b})~\textrm{or}~(\bar{a}+h,\bar{b}+g)~\textrm{and}\\
	(H_2,G_2)=(a,b)~\textrm{or}~(a+h,b+g)\\
\end{array}\right. \\
~\\
\left(\frac{2\eta^3}{\theta[^{1-h}_{1-g}]}\right)^2\bar\theta[^{\bar{a}+H_1}_{\bar{b}+G_1}]^4,~~\textrm{for}
	\left\{\begin{array}{l}
	(h,g)\neq(0,0), (h',g')\neq(0,0) \\
	(h,g)=(h',g')~\textrm{and}\\
	(H_2,G_2)=(a,b)~\textrm{or}~(a+h,b+g)\\
\end{array}\right. \\
~\\
\left(\frac{2\eta^3\cdot 2\bar\eta^3}{\theta[^{1-h}_{1-g}]\bar\theta[^{1-h-h'}_{1-g-g'}]}\right)^2,~~\textrm{for}
	\left\{\begin{array}{l}
	(h,g)\neq(0,0), (h',g')\neq(0,0) \\
	(h,g)\neq(h',g')~\textrm{and}\\
	(H_1,G_1)=(\bar{a},\bar{b})~\textrm{or}~(\bar{a}+h+h',\bar{b}+g+g')~\textrm{and}\\
	(H_2,G_2)=(a,b)~\textrm{or}~(a+h,b+g)\\
\end{array}\right. \\
\end{array}\right.
\ee
The model satisfies the conditions for 
\emph{MSDS} structure discussed in \cite{reducedMSDS}. The partition function at the \emph{MSDS} point in moduli space 
is given by
\be
	Z_{Twisted}(\tau,\bar \tau)=208=\textrm{constant}.
\ee

\subsection{Deformations and stability}

We now address the problem of stability under marginal deformations of the current-current type. 
First, the asymmetric nature of the $\Z_2'$ orbifold projects out all moduli 
associated to the $X^{5,6,7,8}$-directions. 
Indeed, under the general action of the $\Z_2\times\Z_2'$ orbifold, the marginal operators transform as:
\begin{align*}
	J^p(z)\times\bar{J}^q(\bar{z})&\xrightarrow{\Z_2\times\Z_2'} +J^p(z)\times\bar{J}^q(\bar{z}), \\
	J^p(z)\times\bar{J}^J(\bar{z})&\xrightarrow{\Z_2\times\Z_2'}(-)^{g+g'}J^p(z)\times\bar{J}^J(\bar{z}), \\
	J^I(z)\times\bar{J}^q(\bar{z})&\xrightarrow{\Z_2\times\Z_2'}(-)^{g}J^I(z)\times\bar{J}^q(\bar{z}), \\
	J^I(z)\times\bar{J}^J(\bar{z})&\xrightarrow{\Z_2\times\Z_2'}(-)^{g'}J^I(z)\times\bar{J}^J(\bar{z}), \\
\end{align*}
where $p,q=1,2,3,4$ and $I,J=5,6,7,8$.
This has important consequences for the stability of the model. 
The only marginal operators that are invariant under the orbifold and which can, thus, 
be used to perturb the sigma model are those associated entirely with the 
shifted $\Gamma_{(4,4)}^{(1)}$ lattice. The moduli space of the theory is then reduced to
\be
	\frac{SO(8,8)}{SO(8)\times SO(8)} \xrightarrow{\Z_2\times\Z_2'} \frac{SO(4,4)}{SO(4)\times SO(4)}~.
\ee
Note that the special \emph{asymmetric} structure of the orbifold twist projects out all moduli associated to the twisted $\Gamma_{(4,4)}^{(2)}$-lattice and, as a result, \emph{drastically reduces} the rank of the space of propagating moduli.In what follows we show the absence of physical tachyons for any marginal deformation 
in the moduli space of the orbifolded theory. 
For this purpose, the following discussion will be restricted entirely 
to the NS-NS sector, where $a=\bar{a}=0$.

Let us first recall the well-known fact that tachyonic states 
cannot appear in the twisted sectors. 
Indeed, consider the twisted sector corresponding to $h=1$ (the analogous argument holds for $h'$). 
The non-vanishing contribution of the fermions in the $R$-symmetry 
lattice is at least $\theta[^0_1]^2\sim q^{1/4}$, 
whereas the twisted $\Gamma^{(2)}_{(4,4)}$ lattice (twisted bosons) 
contributes (at least chirally) $\left(\frac{2\eta^3}{\theta[^{~0~}_{1\pm 1}]}\right)^2\sim q^{1/4}$. 
Therefore, the twisted spectrum is always at least chirally (or anti-chirally) massless 
and, thus, no tachyonic excitations can appear in these sectors.

We now concentrate our attention on the untwisted NS-NS sector $a=\bar{a}=h=h'=0$, 
coupling to the $O\bar{O}$ fermion characters. 
We will show that the twisted $\Gamma_{(4,4)}^{(2)}$ lattice in this sector 
is always at least chirally (anti-chirally) massless, 
and that the spectrum is free of tachyonic modes, 
for any deformation of the shifted $\Gamma^{(1)}_{(4,4)}$ lattice. 
The twisted lattice has no (untwisted) moduli as a result of 
the asymmetric action of the $\Z_2'$-orbifold, as discussed above. 
It is straightforward to see from (\ref{twistedLattice}) that only the unshifted 
states $H_1=H_2=0$ are in danger\footnote{When we refer to states 
as being in danger of becoming tachyonic, 
we imply that they crucially depend on the mass contribution from the (deformable) 
shifted $\Gamma_{(4,4)}^{(1)}$ lattice in order to maintain a positive squared mass. 
As a result, these states could become tachyonic for some deformation of the shifted lattice.} 
of becoming tachyonic. 
Indeed, any one of the $H_i$-shifts would ``excite''  
the spinorial representation of $SO(4)\times SO(4)$, 
which is always (at least) chirally massless. 
To illustrate the statement, taking $H_2=1$ would imply 
that the non-vanishing contribution has weight:
\be
	\theta[^{~~~1~~~}_{b+G_2+g}]^2 \theta[^{~~1~~}_{b+G_2}]^2 \sim q^{1/2}+\mathcal{O}(q).
\ee
This is also easy to see by noticing that the $H_2$-shifted internal bosons $X^{5,6,7,8}(z)$ 
appear in the spectrum as excitations of the affine primary (spin-)field:
\be
	e^{\frac{i}{2}(\pm X^5\pm X^6\pm X^7 \pm X^8)},
\ee
whose conformal weight for $\Z_2$-shifts is $(1/2,0)$. 
It therefore suffices to check the ``unshifted'' sector $H_1=H_2=0$. 
The twisted lattice (\ref{twistedLattice}) in this sector reads:
\be\label{twistedLattice2}
	\Gamma_{(4,4)}^{(2)}\left.\left[^{0, 0~;~ 0, 0}_{b,\bar{b}~;~ G_1, G_2}\right|{}^{0, 0}_{g, g'}\right] 
= \theta[^{~~~~0~~~~}_{b+G_2+g}]^2 \theta[^{~~0~~}_{b-G_2}]^2~\times~ 
\bar\theta[^{~~~~0~~~~}_{\bar{b}+G_1+g+g'}]^2 \bar\theta[^{~~0~~}_{\bar{b}-G_1}]^2.
\ee
Let us now discuss the GGSO projections that need to be carried out. 
In our formalism, they correspond to summations over $b,\bar{b}, G_1,G_2, g,g'$.
Once the projections are imposed, the above sector of the twisted $\Gamma_{(4,4)}^{(2)}$ 
lattice will be organized in terms of 
$[SO(4)\times SO(4)]_L\times[SO(4)\times SO(4)]_R$ characters in the vectorial $V_4$ ($\bar{V}_4$) 
and vacuum $O_4$ ($\bar{O}_4$) representations, 
whereas the shifted $\Gamma_{(4,4)}^{(1)}$ lattice will be decomposed 
into the $V_8$ ($\bar{V}_8$) and  $O_8$ ($\bar{O}_8$) characters of $SO(8)_L\times SO(8)_R$. 
Of course, for the shifted $\Gamma_{(4,4)}^{(1)}$ lattice this decomposition 
is only valid at the \emph{MSDS} point where the enhanced 
symmetry contains an $SO(8)_L\times SO(8)_R$ factor of global (classification) symmetry.

The subsector of interest is clearly the one in which the left- and right-moving $R$-symmetry 
lattices come only with characters in the vacuum $O, \bar{O}$ representation. 
In this case, the full $\Gamma_{(8,8)}$ lattice has odd $b$ and $\bar{b}$ 
and even $g$, $g'$ GGSO-projections. However, as it is straightforward to see from (\ref{twistedLattice2}), 
imposing an odd projection on the characters of the twisted $\Gamma_{(4,4)}^{(2)}$ lattice 
necessarily leads to chiral (or antichiral) vectorial representations $V_4$ (or $\bar{V}_4$) 
appearing in the spectrum,  and the sector is at least (anti-)chirally massless. 
On the other hand, if we impose even projections on the characters of the twisted lattice, 
the overall odd projections force an odd GGSO-projection on the shifted lattice. 
This subsector is expressed as follows:
\be
	\frac{1}{\eta^8\bar\eta^8}\left.\Gamma_{(8,8)}\right|_{\textrm{subsector}}\rightarrow 
\left(\frac{1}{2^4 \eta^4\bar\eta^4}\sum\limits_{b,\bar{b},G_1, G_2}{(-)^{b+\bar{b}} 
\Gamma_{(4,4)}^{(1)}\left[^{0, 0~;~ 0, 0}_{b,\bar{b}~;~ G_1, G_2}\right]}\right)
~\times~\left(~O_4 O_4\times \bar{O}_4\bar{O}_4~\right),
\ee
where the projections now act only upon the shifted $\Gamma_{(4,4)}^{(1)}$ lattice, 
whereas the twisted $\Gamma_{(4,4)}^{(2)}$ lattice only contributes with 
the vacuum representation in this subsector of interest. 
It is now straightforward to see that this subsector, 
which is the only one dangerous of producing tachyonic modes, 
is in fact projected out by the $b,\bar{b},G_1,G_2$ GGSO projections. 
Indeed, it suffices to consider the above projection at the (undeformed) \emph{MSDS} 
point of moduli space where the simple expression (\ref{shiftedLattice}) 
for the shifted lattice is available. 
Expressed in terms of $SO(2n)$ characters, 
the only state surviving the odd $b,\bar{b}$-projections would be:
\be
	\left(V_8\times \bar{V}_8\right)\times\left(~O_4 O_4\times \bar{O}_4\bar{O}_4~\right).
\ee
However, this state violates the even $G_1, G_2$-projection correlating
the parity of the two lattices and is projected out of the spectrum. 
Since all other sectors are at least (anti-)chirally massless, 
any deformation of the shifted $\Gamma_{(4,4)}^{(1)}$ lattice in $\frac{SO(4,4)}{SO(4)\times SO(4)}$ 
will not produce tachyonic states and the vacuum is, from that point of view, 
stable. 

\subsection{Connection with supersymmetric vacua}

We now discuss various deformation limits and establish 
the continuous connection of the tachyon-free \emph{MSDS} vacua 
with higher-dimensional vacua of Type II superstring theory. 
It will be convenient to shift the summation variables 
$H_1\rightarrow H_1-\bar{a}$, $G_1\rightarrow G_1-\bar{b}$ 
and $H_2\rightarrow H_2-a$, $G_2\rightarrow G_2-b$, 
since they are defined modulo 2. 
With this shifting, the twisted $\Gamma_{(4,4)}$ lattice does not depend on the $R$-symmetry 
charges. The shifted lattice at the fermionic point can be written in the Lagrangian form as:
\be
		\Gamma_{(4,4)}^{(1)}=\sum\limits_{m^i,n^i\in\Z}
{e^{-\frac{\pi}{\tau_2}(G+B)_{ij}(m^i+\tau n^i)(m^j+\bar\tau n^j)+i\pi\mathcal{T}}},
\ee
where $G$ and $B$ were defined in (\ref{metrictorsion}), 
and the phase coupling $\mathcal{T}$ is:
\be
	\mathcal{T}= m^1 n^1+(a+\bar{a}+H_2)m^1+(b+\bar{b}+G_2)n^1
+(H_1+H_2)m^4+(G_1+G_2)n^4+(G_1+G_2)(H_1+H_2).
\ee
It is easy to see that the modulus that couples to the total fermion number $F_L+F_R$ is $G_{11}$, 
corresponding to the current-current deformation $G_{11}\partial X^1\bar\partial X^1$. 
Of course, this is expressed in the transformed basis of Appendix \ref{AppendixLattice}, 
where the lattice frame was repeatedly rotated. 
In terms of the initial bosonization coordinates $X_{(0)}^{1,2,3,4}$, the deformation lies along the ``diagonal'' direction:
\be
	(\partial X_{(0)}^1+\partial X_{(0)}^2+\partial X_{(0)}^3+\partial X_{(0)}^4)
(\bar\partial X_{(0)}^1+\bar\partial X_{(0)}^2+\bar\partial X_{(0)}^3+\bar\partial X_{(0)}^4).
\ee
We now discuss certain interesting points regarding marginal deformations.
\begin{itemize}
\item Taking the infinite $G_{11}\rightarrow\infty$ limit, 
the $X^1$ cycle of $T^4$ decompactifies and the coupling to the $R$-symmetry charges is washed out. 
The resulting model is a supersymmetric $\mathcal{N}=4$ Type II model in $2+1$ spacetime dimensions. 
The new non-compact dimension is interpreted as an emergent spatial-dimension as it can be generated dynamically. 
Furthermore, additional $\Z_2$-orbifold twists could further reduce 
the supersymmetry down to $\mathcal{N}=2$ or $\mathcal{N}=1$, 
without altering the tachyon-free structure of the theory.

\item In the thermal version of the models, Euclidean time is identified with 
the compact toroidal cycle coupled to the total fermion number as in the Hybrid models 
of the previous section. Note that the difference between the two cases, however, 
is that the present breaking of supersymmetry is by no means an arbitrary one. 
It is, rather, \emph{a very specific breaking dictated by the underlying MSDS algebra}. 
Within the thermal interpretation, the moduli associated to the $X^1$ direction 
are interpreted as the thermodynamical parameters of the theory. 
The ``infinite radius" limit $G_{11}\rightarrow\infty$ corresponds 
to zero temperature and yields a supersymmetric 
vacuum with 2 non-compact spatial dimensions. 

\item Other interesting limits involve the radial deformations along the 
3 remaining spatial directions, 
orthogonal to the $X^1$ direction. 
In the infinite (decompactification) limit new non-compact dimensions emerge continuously. 
In particular, one may obtain three-dimensional vacua, 
with supersymmetry spontaneously broken by thermal or gravito-magnetic fluxes or, 
in the infinite limit, four-dimensional superstring vacua 
(superstring vacua with 3 non-compact spatial dimensions at zero temperature, in the thermal version).

\end{itemize}

%%%%%%%%%%%%%%%%%%%%%%%%%%%%%%%%%%%%%%%%%%%%%%%%%%%%%%%%%%%%%%%%%%%%%%%%%%%
%%%%%%%%%%%%%%%%%%%%%%%%%%%%%%%%%%%%%%%
\section{Conclusions}

In this paper we studied current-current type marginal deformations of \emph{MSDS} vacua \cite{massivesusy} and their orbifolds \cite{reducedMSDS}. We explicitly identified the relevant half-shifted  $\Gamma_{(8,8)}[^{\,a\,,\,\bar{a}\,}_{\,b\,,\,\bar{b}\,}]$ lattices, which exhibit the maximally symmetric \emph{MSDS} models as special points in the moduli space of Type II and Heterotic compactifications to two dimensions. It was shown that \emph{MSDS} models admit at most two independent moduli, which participate in the supersymmetry breaking via couplings to the left- and right-moving spacetime fermion numbers $F_L,\, F_R$, respectively. 
We showed the existence of marginal deformations interpolating between \emph{MSDS} vacua (and their orbifolds) and conventional four-dimensional string models where supersymmetry is spontaneously broken by geometrical fluxes. This establishes a correspondence between the \emph{MSDS} space of vacua and four-dimensional gauged supergravities of the ``no-scale" type  \cite{Noscale}.  

We would like to stress here, that this correspondence is of fundamental importance within the string cosmological framework, where the \emph{MSDS}-vacua, free of tachyons and other pathological instabilities, become natural candidates to describe the very early stringy, ``non-geometric era'' of the Universe. This is because the correspondence permits one to connect, at least adiabatically, the initially two-dimensional $MSDS$ vacua with \emph{semi-realistic} four-dimensional string vacua, 
where spacetime supersymmetry is spontaneously broken at late cosmological times via thermal or supersymmetry breaking moduli. 
Specifically, the thermal interpretation of various Euclidean versions of these exotic constructions was analysed in detail and the finite-temperature description was presented in terms of thermal ensembles, further deformed by discrete gravito-magnetic fluxes. 

Furthermore, the stability of \emph{MSDS} vacua under arbitrary marginal deformations was extensively analysed, allowing the identification of deformation directions along which tachyonic instabilities are generically encountered. 
An important result of this paper is that such ``dangerous'' deformation moduli can be projected out by introducing asymmetric $\Z_2\times \Z_2$ orbifolds. Such tachyon-free orbifolds are not only compatible with the \emph{MSDS} structure but also reduce the number of (spontaneously broken) supersymmetries restored at late cosmological times. The necessary stability conditions, under arbitrary deformations of the dynamical moduli, were determined for general (untwisted) thermal Type II \emph{MSDS} vacua. Once met, the corresponding ensemble is equivalent to the right-moving spacetime fermion index and the model is characterized by thermal duality symmetry.    

A very special class of \emph{Hybrid MSDS} models, sharing similar properties as those described in the above correspondence, were constructed in this work. These two-dimensional Type II models are characterized by an asymmetry in the left- and right- moving sectors of the theory. Namely, the left-moving side enjoys conventional supersymmetry (which can be regarded as a special case of \emph{MSDS} symmetry), whereas the right-moving side realizes the \emph{MSDS} structure. These models were shown to admit a natural implementation of temperature, where the left-moving supersymmetries are spontaneously broken by thermal effects.
The thermal partition function of the models was analysed and was found to be finite for all values of the temperature, and to be characterized by thermal duality symmetry. The two dual phases were shown to be connected via a first-order phase transition at the \emph{MSDS} point. In the limit where some of the spatial dimensions decompactify, the phase transition becomes milder.

Obviously, the adiabatic connection outlined above is not sufficient to define the complete cosmological evolution from the early \emph{MSDS} era towards the late time Universe, described by $\mathcal{N}=1$ spontaneously broken supersymmetry. A dynamical realization of the adiabatic evolution is still lacking, especially during the non-geometrical era close to the extended symmetry (\emph{MSDS}) points, where conventional effective field-theoretic techniques are still absent. Once some of the moduli become sufficiently large, so that a conventional spacetime description emerges, the subsequent evolution in the intermediate cosmological regime can be unambiguously described, as shown in Ref.\cite{ckpt,CriticalCosmo} (see also \cite{ak} for related earlier work). There, an attractor mechanism was discovered dominating the late-time cosmological evolution, depending only on the particular structure of supersymmetry breaking induced by the fluxes.  The fact that the initial \emph{MSDS} structure unambiguously determines these fluxes, strongly indicates that we are in a good direction. Furthermore, the qualitative infrared behavior of string-induced effective gauged supergravity suggests that we are describing a ``non-singular string evolutionary scenario'' connecting particle physics and cosmology.

The above attractive scenario crucially depends on the spontaneous dynamical exit from the early non-geometrical \emph{MSDS} phase. In this respect, it is possible to construct (at least Heterotic) \emph{MSDS} vacua 
whose massless spectra are characterized by an abundance of fermionic (rather than bosonic) degrees of freedom $n_F>n_B$ \cite{MoreFermions}. There are strong indications that this configuration induces 
a quantum instability (at the one-loop level) that could trigger the desired cosmological evolution,
providing the spontaneous exit from the early \emph{MSDS} era. The details of this investigation will be considered elsewhere \cite{MoreFermions}.
 
Finally, let us conclude by restating that the problem of initial \emph{MSDS}-vacuum selection is a highly non-trivial one, because of constraints arising from the observed particle phenomenology at late cosmological times. Demanding that late-time four-dimensional semi-realistic ${\cal N}= 1$ string vacua, with phenomenologically viable gauge interactions (such as those based on an $SO(10)$ GUT-gauge group \cite{FarKRizos}) are in adiabatic correspondence with the initial \emph{MSDS}-vacua, impose severe restrictions on the initial vacuum space. A study of these restrictions is expected to significantly reduce the number of initial candidate vacua and is currently under investigation \cite{chiralhet}.

%%%%%%%%%%%%%%%%%%%%%%%%%%%%%%%%%%%%
\section*{Acknowledgements}

We are grateful to C. Bachas, A. Faraggi, H. Partouche, J. Rizos and J. Troost for useful discussions.
N.T. thanks the Ecole Normale Sup\'{e}rieure, I.F. and C.K. thank the University of Cyprus for hospitality during part of this work. This work is partially supported by the IFCPAR programme 4104-2 and the ANR programme NT09-573739.
%%%%%%%%%%%%%%%%%%%%%%%%%%%%%%%%%%%%

\appendix
\numberwithin{equation}{section}

\section{The half-shifted $(8,8)$-lattice}\label{AppendixLattice}

In this appendix we give a detailed derivation of the form of the half-shifted $(8,8)$ lattice 
of equation (\ref{FinalLatt}).
We begin with the lattice in the form of equation (\ref{latt2}),  
and split the lattice directions into two groups of four, indicated by
primed and unprimed quantum numbers, respectively, and define
$$
	v_i \equiv m_i + \tau n_i,
$$
$$
	w_i \equiv m'_i+\tau n'_i,
$$
$$
	V\equiv M+\tau N,~~W\equiv M'+\tau N',
$$
\be
	\xi\equiv g+\tau h,
\ee
where $i=1,2,3,4$, and
$$
M\equiv \sum\limits_{i=1}^4{m_i}~~N\equiv\sum\limits_{i=1}^4{n_i},
$$
\be
M'\equiv \sum\limits_{i=1}^4{m'_i}~~N'\equiv\sum\limits_{i=1}^4{n'_i}~.
\ee
In terms of these variables, the $(8,8)$ lattice (\ref{latt2}) factorizes into two $(4,4)$ sub-lattices. 
The exponent of the first $(4,4)$ sub-lattice is given by
\be\nonumber
	-\frac{\pi}{8\tau_2}\left(|V+2\xi|^2+|V-2v_3-2v_4|^2+|V-2v_2-2v_3|^2+|V-2v_2-2v_4|^2\right)
\ee
\be\nonumber
	+i\pi\left[MN+\left(a+\frac{h}{2}\right)M+\left(b-\frac{g}{2}\right)N+M(n_2+n_3+n_4)-N(m_2+m_3+m_4)\right.
\ee
\be	
	\left.\phantom{\frac{X}{X}}+(m_2 n_3-m_3 n_2)+(m_2+m_3)n_4-m_4(n_2+n_3)\right],
\ee
which is the representation of the Cartan matrix of $E_8$. 
The phase couplings $m_I n_J$ with $I<J$ correspond to the parallelized torsion $B_{IJ}$. 
A similar expression holds for the $(4,4)$ lattice spanned by the primed quantum numbers.

We next shift the summation variables in the following order: 
$V\rightarrow V-2\xi$, $v_3\rightarrow v_3-v_4-\xi$ and $v_2\rightarrow v_2-v_3+v_4$, 
and the lattice exponent becomes
\be
	-\frac{\pi}{8\tau_2}\left(|V|^2+|V-2v_3|^2+|V-2v_2|^2+|-V+2\xi+4v_4+2v_2-2v_3|^2\right)
\ee
along with the phase
\be\nonumber
	i\pi\left[MN+\left(a-\frac{h}{2}\right)M+\left(b+\frac{g}{2}\right)N-M(n_2+n_4)+(m_2+m_4)N\right.
\ee
\be
	\left.\phantom{\frac{X}{X}}+(m_2+m_3)h+(n_2+n_3)g-(m_2 n_3-m_3 n_2)\right].
\ee
Performing a double Poisson re-summation on $v_4$ (see Appendix \ref{DoublePoisson}), 
which takes the corresponding torus to its dual picture ($\tilde{m}_4\equiv n_4$, $\tilde{n}_4\equiv -m_4$), 
allows for the $(h,g)$-dependence to appear only in the phase:
$$
	-\frac{\pi}{8\tau_2}\left( |V|^2+|V-2v_2|^2+|V-2v_3|^2+|V+2\tilde{v}_4-2v_2+2v_3|^2\right)
$$
$$
	+i\pi\left[MN+\bar{a}M+\bar{b}N+(a+\bar{a})\tilde{m}_4+(b+\bar{b})\tilde{n}_4 \phantom{\frac{X}{X}}\right.
$$
\be\label{master1}
	\left.+\frac{1}{2}M\tilde{n}_4-\frac{1}{2}\tilde{m}_4N-(Mn_2-m_2N)-(m_2 n_3-m_3 n_2)-
(m_2 \tilde{n}_4-\tilde{m}_4 n_2) +(m_3\tilde{n}_4-\tilde{m}_4 n_3)\right].
\ee
From this expression of the $(4,4)$ sub-lattice, we obtain the metric $G$ and parallelized torsion $B$:
\be\label{metrictorsion}
	G= \left(\begin{array}{r r r r}
	\frac{1}{2}&-\frac{1}{2}&0&\frac{1}{4}\\
 -\frac{1}{2}& 1&-\frac{1}{2}&-\frac{1}{2}\\
 0&-\frac{1}{2}&1&\frac{1}{2}\\
 \frac{1}{4}&-\frac{1}{2}&\frac{1}{2}&\frac{1}{2}\\
\end{array}\right)~,~
B=\left(\begin{array}{r r r r}
	0&-\frac{1}{2}&0&\frac{1}{4}\\
 \frac{1}{2}& 0&-\frac{1}{2}&-\frac{1}{2}\\
 0&\frac{1}{2}&0&\frac{1}{2}\\
 -\frac{1}{4}&\frac{1}{2}&-\frac{1}{2}&0\\
\end{array}\right).
\ee
Notice that the sum $E\equiv G+B$ is an upper triangular matrix, 
as required for the holomorphic factorization of the theory 
and the presence of a Kac-Moody algebra of extended symmetry.

Adding the contribution of the remaining $(4,4)$-sub-lattice, we obtain the full $(8,8)$ lattice:
$$
-\frac{\pi}{8\tau_2}\left( |V|^2+|V-2v_2|^2+|V-2v_3|^2+|V+2\tilde{v}_4-2v_2+2v_3|^2\right.
$$
$$
\left. +|W|^2+|W-2w_2|^2+|W-2w_3|^2+|W+2\tilde{w}_4-2w_2+2w_3|^2\right)
$$
$$
	+i\pi\left[MN+M'N'+\bar{a}(M+M')+\bar{b}(N+N')
+(a+\bar{a})(\tilde{m}_4+\tilde{m}'_4)+(b+\bar{b})(\tilde{n}_4+\tilde{n}'_4) \phantom{\frac{X}{X}}\right.
$$
$$
	\left.+\frac{1}{2}M\tilde{n}_4-\frac{1}{2}\tilde{m}_4N-(Mn_2-m_2N)
-(m_2 n_3-m_3 n_2)-(m_2 \tilde{n}_4-\tilde{m}_4 n_2) +(m_3\tilde{n}_4-\tilde{m}_4 n_3)\right.
$$
\be
	\left.+\frac{1}{2}M'\tilde{n}'_4-\frac{1}{2}\tilde{m}'_4N'-(M'n'_2-m'_2N')
-(m'_2 n'_3-m'_3 n'_2)-(m'_2 \tilde{n}'_4-\tilde{m}'_4 n'_2) +(m'_3\tilde{n}'_4-\tilde{m}'_4 n'_3)\right].
\ee
The full metric and torsion are $8\times 8$ symmetric and antisymmetric matrices 
expressed in block-diagonal form in terms of the $4\times 4$ matrices $G$ and $B$ of (\ref{metrictorsion}):
\be
	\textbf{G}=\left(
\begin{array}{c c}
	G & 0 \\
	0 & G \\
\end{array}\right)~~,~~\textbf{B}=\left(\begin{array}{c c}
	B & 0 \\
	0 & B \\
\end{array}\right).
\ee
The source of supersymmetry breaking in the Type II models is the coupling to the momentum lattice of the two independent $R$-symmetry charges:
the left- and right- moving spacetime fermion numbers $F_L$ and $F_R$. It is then clear that one needs 
to infinitely deform at most two independent ``radial'' moduli in order 
to recover a maximally supersymmetric vacuum which, in fact, will be four-dimensional. 
To identify these, we cast the coupling in left/right symmetric form \cite{akpt}, 
by shifting $(V,W)\rightarrow (V-\tilde{v}_4,W-\tilde{w}_4)$:
\be\nonumber
	-\frac{\pi}{8\tau_2}\left(|V-\tilde{v}_4|^2
+|V-\tilde{v}_4-2v_2|^2+|V-\tilde{v}_4-2v_3|^2+|V+\tilde{v}_4-2v_2+2v_3|^2 \right.
\ee
\be\nonumber
	\left.   |W-\tilde{w}_4|^2+|W-\tilde{w}_4-2w_2|^2+|W-\tilde{w}_4-2w_3|^2+|W+\tilde{w}_4-2w_2+2w_3|^2 \right)
\ee
\be\nonumber
	+i\pi\left( MN+M'N'+\tilde{m}_4 \tilde{n}_4+\tilde{m}'_4 \tilde{n}'_4
+a(\tilde{m}_4+\tilde{m}'_4)+b(\tilde{n}_4+\tilde{n}'_4)+\bar{a}(M+M')+\bar{b}(N+N')\right.
\ee
\be
	\left.+2B_{IJ}m_I n_J\right).
\ee
By further shifting $V\rightarrow V-W$, $\tilde{w}_4\rightarrow \tilde{w}_4-\tilde{v}_4$, 
we can completely isolate the $R$-symmetry couplings to within a $(2,2)$ sub-lattice:
\be\nonumber
	-\frac{\pi}{8\tau_2}\left(|V-W-\tilde{v}_4|^2+|V-W-\tilde{v}_4-2v_2|^2
+|V-W-\tilde{v}_4-2v_3|^2+|V-W+\tilde{v}_4-2v_2+2v_3|^2 \right.
\ee
\be\nonumber
	\left.   |W-\tilde{w}_4+\tilde{v}_4|^2+|W-\tilde{w}_4+\tilde{v}_4-2w_2|^2
+|W-\tilde{w}_4+\tilde{v}_4-2w_3|^2+|W+\tilde{w}_4-\tilde{v}_4-2w_2+2w_3|^2 \right)
\ee
\be\label{NewLattice}
	+i\pi\left(  MN+\bar{a}M+\bar{b}N 
+\tilde{m}'_4\tilde{n}'_4+a\,\tilde{m}'_4+b\,\tilde{n}'_4+ 2B_{IJ}m_I n_J \right).
\ee
As in \cite{akpt}, one cycle of this $(2,2)$ lattice is ``thermally'' coupled to $F_L$ and the other is ``thermally''
coupled to $F_R$.

Finally, it is convenient to use the reshuffled basis 
$\textbf{m}^T=(M,\tilde{m}'_4|m_2,m_3,\tilde{m}_4,M',m'_2,m'_3)$,  
$\textbf{n}^T=(N,\tilde{n}'_4|n_2,n_3,\tilde{n}_4,N',n'_2,n'_3)$. 
In the new basis the metric $G_{IJ}$ and antisymmetric tensor $B_{IJ}$ are:
\be\label{metric}
	G_{IJ}=\left(\begin{array}{r r | r r r r r r}
	\frac{1}{2} & 0 & -\frac{1}{2} & 0 & -\frac{1}{4} & -\frac{1}{2} & 0 & 0 \\
	0 & \frac{1}{2} & 0 & 0 & -\frac{1}{2} & -\frac{1}{4} & 0 & \frac{1}{2} \\ \hline
	-\frac{1}{2} & 0 & 1 & -\frac{1}{2} & 0 & \frac{1}{2} & 0  & 0 \\
	0 & 0 & -\frac{1}{2} & 1 & \frac{1}{2} & 0 & 0 & 0 \\
	-\frac{1}{4} & -\frac{1}{2} & 0 & \frac{1}{2} & 1 & \frac{1}{2} & 0 & -\frac{1}{2}\\
	-\frac{1}{2} & -\frac{1}{4} & \frac{1}{2} & 0 & \frac{1}{2} & 1 & -\frac{1}{2} & 0 \\
	0 & 0 & 0 & 0 & 0 & -\frac{1}{2} & 1 & -\frac{1}{2} \\
	0 & \frac{1}{2} & 0 & 0 & -\frac{1}{2} & 0 & -\frac{1}{2} & 1\\
\end{array}\right)
\ee

\be\label{torsion}
	B_{IJ}=\left(\begin{array}{r r | r r r r r r}
	0 & 0 & -\frac{1}{2} & 0 & -\frac{1}{4} & \frac{1}{2} & 0 & 0 \\
	0 & 0 & 0 & 0 & \frac{1}{2} & \frac{1}{4} & 0 & -\frac{1}{2} \\ \hline
	\frac{1}{2} & 0 & 0 & -\frac{1}{2} & 0 & \frac{1}{2} & 0  & 0 \\
	0 & 0 & \frac{1}{2} & 0 & \frac{1}{2} & 0 & 0 & 0 \\
	\frac{1}{4} & -\frac{1}{2} & 0 & -\frac{1}{2} & 0 & -\frac{1}{2} & 0 & -\frac{1}{2}\\
	-\frac{1}{2} & -\frac{1}{4} & -\frac{1}{2} & 0 & \frac{1}{2} & 0 & -\frac{1}{2} & 0 \\
	0 & 0 & 0 & 0 & 0 & \frac{1}{2} & 0 & -\frac{1}{2} \\
	0 & \frac{1}{2} & 0 & 0 & \frac{1}{2} & 0 & \frac{1}{2} & 0\\
\end{array}\right),
\ee
with $I=1,2,\ldots 8$.

%%%%%%%%%%%%%%%%%%%%%%%%%%%%%%%%%%%%%%%%%%%%%%%%%%%

\section{Double Poisson Resummation}\label{DoublePoisson}

A useful formula that takes a lattice direction to its dual representation is obtained by double Poisson re-summation over 
both winding numbers $m,n$ (defined in the Lagrangian representation): 
\be\label{DoublePoissonForm}
\frac{R}{\sqrt{\tau_2}}\sum\limits_{m,n\in\mathbb{Z}}{e^{-\frac{\pi R^2}{\tau_2}\left|m+\frac{g}{2}+\tau\left(n+\frac{h}{2}\right)\right|^2+i\pi\left(mA+nB\right)}} = \frac{1/R}{\sqrt{\tau_2}}\sum\limits_{m,n\in\mathbb{Z}}{e^{-\frac{\pi(1/R)^2}{\tau_2}\left|n-\frac{B}{2}-\tau\left(m-\frac{A}{2}\right)\right|^2+i\pi\left[h\left(n-\frac{B}{2}\right)+g\left(m-\frac{A}{2}\right)\right]}}.
\ee
It is a special case  of the general duality transformations of the $O(d,d;\Z)$-duality group.

%%%%%%%%%%%%%%%%%%%%%%%%%%%%%%%%%%%%%%%%%%%%%%%%%%%%

\section{Mass Spectrum Generalities}

We briefly state the lattice contribution to the mass spectrum of the deformed vacua discussed in the text. The $(8,8)$-lattices under consideration can be written in Lagrangian form as:
\be
	\Gamma_{(8,8)}=\frac{\det{G}}{(\sqrt{\tau_2})^8}\sum\limits_{m_i,n_i\in\Z}{ e^{-\frac{\pi}{\tau_2}(G+B)_{IJ}(m_I+\tau n_I)(m_J+\bar\tau n_J)+i\pi\mathcal{T}} },
\ee
where $\mathcal{T}$ is a phase coupling that may, in general, break supersymmetry. We parametrize it as follows:
\be
\mathcal{T} = \sum\limits_I{C_I(m_I n_I + a_I m_I + b_I n_I)} + D_I m_I + E_I n_I.
\ee
The form of this coupling is dictated by modular invariance. Here, $a_I$ and $b_I$ are generalized spin structures that may break supersymmetry.
The constants $D_I$ and $E_I$ can be seen as column vectors in the space of $(m_I, n_I)$, whereas $C_I$ can be taken to be diagonal matrices. In the latter notation, the phase can be rewritten as:
\be
	\mathcal{T}= \textbf{m}^T(\textbf{C}\textbf{n}+\textbf{D}+\textbf{C}\textbf{a})+\textbf{b}^T \textbf{C} \textbf{n}+\textbf{E}^T \textbf{n}.
\ee
The last two terms do not participate in the Poisson resummation with respect to $m_I$. However, they are important for the implementation of  GGSO-projections\footnote{In general, $D_I$ and $E_I$ may depend on the spin structures $a_I$, $b_I$.}. Let us define the vector:
\be
	\textbf{S} \equiv \textbf{C}\textbf{n}+\textbf{D}+\textbf{C}\textbf{a},
\ee
which shifts all momenta by half a unit. A standard Poisson re-summation on the momenta $m_I$ will give us the usual mass formulae for toroidal compactifications shifted by $\textbf{S}$:
\be
	P_L^2 = \frac{1}{2}\left(\textbf{m}-\frac{1}{2}\textbf{S}+(\textbf{B}+\textbf{G})\textbf{n}\right)^T\textbf{G}^{-1}\left(\textbf{m}-\frac{1}{2}\textbf{S}+(\textbf{B}+\textbf{G})\textbf{n}\right)
\ee
\be
	P_R^2 = \frac{1}{2}\left(\textbf{m}-\frac{1}{2}\textbf{S}+(\textbf{B}-\textbf{G})\textbf{n}\right)^T\textbf{G}^{-1}\left(\textbf{m}-\frac{1}{2}\textbf{S}+(\textbf{B}-\textbf{G})\textbf{n}\right)
\ee
where the lattice is now written in its Hamiltonian form:
\be
	\Gamma_{(8,8)}=\sum\limits_{P_L,P_R\in\Lambda}{ q^{\frac{1}{2}P_L^2}\,\bar{q}^{\frac{1}{2}P_R^2} } .
\ee
The level-matching contribution of the lattice then reads:
\be\label{levelmatching}
	\frac{1}{2}(P_L^2-P_R^2) = \textbf{m}^T\textbf{n}-\frac{1}{2}\,\textbf{S}^T\textbf{n},
\ee
and is, of course, independent of the deformation parameters.

%%%%%%%%%%%%%%%%%%%%%%%%%%%%%%%%%%%%%%%%%%%%%%%

\section{Conditions for Tachyon Finiteness: a technical point}\label{conditionsCheck}

In this Appendix we demonstrate that the conditions for tachyon finiteness derived in Section \ref{stability} are equivalent to the factorization of a $\Gamma_{(1,1)}$-lattice coupled \emph{chirally} to the left- (or right-) moving $R$-symmetry charges. This result, in turn, implies that the only\footnote{Again, as in the main body of the paper, we implicitly assume that all moduli are propagating fields, except for the ones associated to the Euclidean time direction, which acquire a thermodynamical interpretation. In the case where some of the moduli are twisted by orbifolds, as in Section \ref{twisted}, these conditions are no longer applicable in this simplified form.} tachyon-free thermal vacua are the ones where the temperature cycle $S^1$ is completely factorized from the compact manifold.  This is essentially equivalent to showing that, by a discrete $O(8,\mathbb{Z}){\times}O(8,\mathbb{Z})\in O(8,8;\mathbb{Z})$-rotation, the stability conditions (\ref{TachyonFreeCond}) can always be brought to the form $\hat{G}_0^1=2\hat{B}_{01}=\pm 1$ and $\hat{G}_0^k=\hat{B}_{0k}=0$, for $k=2,\ldots 7$. 

We begin by parametrizing the $d$-dimensional dual vielbein $(e^{*,a})^\mu\equiv (E^{T,-1}_{(d)})_{a\mu}$ in the (dual) lattice frame in terms of a \emph{unique\footnote{The uniqueness of the decomposition of a symmetric, positive-definite matrix into the product of an upper- and lower- triangular matrix is known in linear algebra as the Cholesky theorem.} upper-triangular} $d\times d$-square matrix:
\begin{align}
		(E^{T,-1}_{(d)})_{\mu\nu} = 
%		\left( 
%\begin{array}{c | c c c}
%	R_0^{-1} & R_0^{-1} G_{01} & \ldots & R_0^{-1} G_{0d} \\ \hline
%	0 & R_1^{-1} & \ldots & R_1^{-1}G_{1d} \\
%	\vdots & \vdots & \ldots & \vdots \\
%	0 & 0 & \ldots & R_d^{-1} \\
%\end{array}\right)= 
\left( \begin{array}{c | c}
	R_0^{-1} & -R_0^{-1} \hat{G}_{0}^{I}  \\ \hline
	0 & (E^{-1,T}_{(d-1)})_{IJ}  \\
\end{array}\right),
\end{align}
where the greek indices $\mu=0,1,\ldots d-1$ run over the entire $d$-dimensional space, while the latin indices span the $d-1$-dimensional subspace $I=1,\ldots d-1$. We also used the $(d-1)$-dimensional dual upper-triangular vielbein matrix $(E^{T,-1}_{(d-1)})_{IJ}$, parametrizing the lattice metric of the $(d-1)$-dimensional subspace spanned by $X^I$, $I=1,\ldots d-1$. 

Note that the first $\mu=0$ row of this matrix contains the $\hat{G}_{0}^{I}$-quantities, identified as chemical potentials in the case of the $(8,8)$-lattice (Section \ref{stability}):
$$
	(e^{*,a=0})^I = - R_0\hat{G}_0^I .
$$

In the Lagrangian representation of the lattice, the exponent can be expressed in terms of the lower-triangular vielbein matrix $(e^a)_\mu\equiv (E_{(d)})_{a\mu}$ as:
\begin{align}
	-\frac{\pi}{\tau_2}~G_{\mu\nu}(\tilde{m}^\mu+\tau n^\mu)(\tilde{m}^\nu+\bar{\tau}n^\nu) = -\frac{\pi}{\tau_2}~\sum\limits_{\mu=0}^{d-1}{\left| \sum\limits_{\nu=0}^{\mu}{(E_{(d)})_{\mu\nu}\, v^\nu}\right|^2},
\end{align}
where we use the notation $v^\mu\equiv \tilde{m}^\mu+\tau n^\mu$, introduced in Appendix \ref{AppendixLattice}. Using standard blockwise inversion, one may express the vielbein matrix as:
\begin{align}
	(E_{(d)})_{\mu\nu} = \left( \begin{array}{c | c}
	R_0 & 0  \\ \hline
	(E_{(d-1)})_{IK}\hat{G}_{0}^{K} & (E_{(d-1)})_{IJ}  \\
\end{array}\right),
\end{align}
so that the lattice exponent has the simple decomposition:
\begin{align}
	-\frac{\pi}{\tau_2}~\left(R_0^2|v^0|^2+\sum\limits_{I=1}^{d-1}{\left| \sum\limits_{J=1}^{I}{(E_{(d-1)})_{IJ}\, \left(v^J+\hat{G}_0^J v^0 \right)}\right|^2}\right).
\end{align}
Now, the condition $\hat{G}_0^{I}\in\mathbb{Z}$ permits the change of lattice basis (discrete gauge transformation):
\begin{align}\label{changeBasis}
	v^I\rightarrow {v'}^{I}=v^I+\hat{G}_0^{I} v^0,
\end{align}
after which, the exponent of the lattice takes a factorized form:
\begin{align}\label{exponentFact}
	-\frac{\pi}{\tau_2}~\left(R_0^2|v^0|^2+\sum\limits_{I=1}^{d-1}{\left| \sum\limits_{J=1}^{I}{(E_{(d-1)})_{IJ}\, {v'}^J  }\right|^2}\right).
\end{align}
We must also ensure that the same factorization takes place in the phase. Working in the ``temperature representation" (\ref{temperatureRep}), the phase:
\begin{align}
	i\pi\left[ \tilde{m}^0(a+\bar{a}) + n^0(b+\bar{b}) + \tilde{m}^1 n^1 + \tilde{m}^1\bar{a}+n^1\bar{b} + 2\sum\limits_{\mu<\nu}^{d}{ B_{\mu\nu}\left(\tilde{m}^{\mu} n^{\nu}-\tilde{m}^\nu n^\mu\right)} \right],
\end{align}
transforms under the change of basis (\ref{changeBasis}). The antisymmetric tensor part transforms as:
\begin{align}
	2\sum\limits_{\mu<\nu}^{d}{B_{\mu\nu}\left(\tilde{m}^\mu n^\nu-\tilde{m}^\nu n^\mu\right)} = 2\sum\limits_{J=1}^{d}{\hat{B}_{0J}\left(\tilde{m}^0 {n'}^{J}-\tilde{m}'\phantom{}^{J} n^0\right)} + 2\sum\limits_{I<J}^{d}{B_{IJ}\left(\tilde{m}'\phantom{}^I {n'}^J-\tilde{m}'\phantom{}^{J} {n'}^I\right)}.
\end{align}
The condition $\hat{B}_{0k}\in\mathbb{Z}$ for $k=2,\ldots d$, eliminates the associated $\hat{B}_{0k}$-term from the phase above, while the $\hat{B}_{01}\in \mathbb{Z}+\frac{1}{2}$-condition ensures that the $\hat{B}_{01}$-term cancels the extra phase contribution of the right-moving $R$-symmetry coupling. As a result, the phase factorizes as well:
\begin{align}\label{phaseFact}
	i\pi\left[ \left(\tilde{m}^0 n^0+ \tilde{m}^0 a + n^0 b\right) + \left(\tilde{m}'\phantom{}^{1} {n'}^{1} + \tilde{m}'\phantom{}^{1}\bar{a}+{n'}^{1}\bar{b}\right) + 2\sum\limits_{I<J}^{d}{ B_{IJ}\left(\tilde{m}'\phantom{}^{I} {n'}^{J}-\tilde{m}'\phantom{}^{J} {n'}^{I}\right)} \right].
\end{align}
The conditions (\ref{TachyonFreeCond}), therefore, imply that the original $(d,d)$-lattice \emph{factors out} a $\Gamma_{(1,1)}[^a_b]$-lattice coupled only to $F_L$:
\begin{align}
	\Gamma_{(d,d)}[^{a,\bar{a}}_{b,\bar{b}}] \rightarrow \Gamma_{(1,1)}[^a_b](R_0) \cdot \Gamma_{(d-1,d-1)}[^{\bar{a}}_{\bar{b}}](G_{IJ},B_{IJ}),
\end{align}
which is, indeed, equivalent to the complete factorization of the temperature cycle $S^1$ from the remaining compact manifold.

%%%%%%%%%%%%%%%%%%%%%%%%%%%%%%%%%%%%%%%%%%%%%%%

\section{Duality}\label{Duality}
It is important to note the presence of an $R_0\rightarrow 1/(2R_0)$ duality as a consequence of the properties of the $\Gamma_{(1,1)}[^a_b]$ shifted lattice. The duality can be seen as follows: changing $\tilde{m_0}, n_0$ variables into $\tilde{m_0}=2M+g$, $n_0=2N+h$, where $M,N\in\Z$ and $h,g\in\{0,1\}$, the lattice takes the form of a $(M+\frac{g}{2},n+\frac{h}{2})$ half-shifted lattice in the presence of a discrete Wilson line co-cycle $ag+bh+hg$:
\be
	\Gamma[^a_b](R_0)=\frac{R_0}{\sqrt{\tau_2}}\sum\limits_{M,N\in\Z}\sum\limits_{h,g=0,1}{e^{-\frac{\pi (2R_0)^2}{\tau_2}\left|\left(M+\frac{g}{2}\right)+\tau\left(N+\frac{h}{2}\right)\right|^2 +i\pi[\,ag+bh+hg\,]}}.
\ee
Next, perform a double-Poisson re-summation as in the Appendix:
\be
	\frac{1}{2^2}\sum\limits_{h,g=0,1}\frac{1/R_0}{\sqrt{\tau_2}}\sum\limits_{M,N\in\Z}{e^{-\frac{\pi}{\tau_2}\left(\frac{1}{2R_0}\right)^2 |N+\tau M|^2+i\pi[MN+Ma+Nb+(M+b+g)(N+h+a)]}(-)^{ab}}.
\ee
Next, one performs another change of variables into $N=2K+G$, $M=2L+H$, where the summations are again over $K,L\in\Z$ and $H,G=0,1$. The result looks very similar to the original form:
\be
	\frac{1}{2}\sum\limits_{h,g,H,G=0,1}\frac{(2R_0)^{-1}}{\sqrt{\tau_2}}\sum\limits_{K,L\in\Z}{e^{-\frac{\pi(1/R_0)^2}{\tau_2} \left|\left(K+\frac{G}{2}\right)+\tau\left(L+\frac{H}{2}\right)\right|^2+i\pi[aG+bH+HG+(G+b+g)(H+h+a)]}(-)^{ab}}.
\ee
The summation over $g$ acts as a projector:
$$
	\sum\limits_{h=0,1}\frac{1}{2}\sum\limits_{g=0,1}{ (-)^{g(H+h+a)}(-)^{(G+b)(H+h+a)}}.
$$
The $g$-projection imposes $H+h+a=\textrm{even}$, whereas the $h$-summation resets $H$ to be unconstrained $H=0,1$. The result is:
$$
	\frac{R_0'}{\sqrt{\tau_2}}\sum\limits_{K,L\in\Z}\sum\limits_{H,G}{e^{-\frac{\pi{R_0'}^2}{\tau_2}|(2K+G)+\tau(2L+H)|^2+i\pi[(2K+G)a+(2L+H)b+(2K+G)(2L+H)+ab]}}\equiv (-)^{ab}\Gamma_{(1,1)}[^a_b](R_0'),
$$
which is exactly the original lattice at the dual radius $R_0'=(2R_0)^{-1}$ and $(-)^{ab}$ is the phase that interchanges the $S$ and $C$ representations.

%%%%%%%%%%%%%%%%%%%%%%%%%%%%%%%%%%%%%%%%%
%%%%%%%%%%%%%%%%%%%%%%%%%%%%%%%%%%%%%%%%%

\end{document}